\documentclass[12pt]{JHEP3}
\pdfoutput=1
\usepackage{amsfonts, amsmath, amsthm}
\usepackage{graphicx}
\title{Lessons from giant gravitons on $AdS_{5}\times T^{1,1}$}
\author{Alex Hamilton$^{1}$\footnote{hamilton@nassp.uct.ac.za}, Jeff Murugan$^{1,2}$\footnote{jeff@nassp.uct.ac.za} and
Andrea Prinsloo$^{1}$\footnote{andrea.prinsloo@uct.ac.za} \\

$^{1}$Cosmology and Gravity Group, \\
Department of Mathematics and Applied Mathematics, \\
University of Cape Town, \\
Private Bag, Rondebosch, 7700, \\
South Africa.\\

$^{2}$National Institute for Theoretical Physics (NITHeP), \\
Private Bag X1, \\
Matieland, 7602, \\
South Africa.}

\abstract{We implement Mikhailov's holomorphic curve construction to explore various properties of giant gravitons in type IIB string theory on $AdS_{5}\times T^{1,1}$.  By coloring the D-brane worldvolume, we are able to show how, in the string theory, the giant graviton factorizes at its maximal size into two dibaryons - topologically stable D-branes wrapping non-contractible cycles in the $T^{1,1}$. This is related to the structure of the symmetry group of the emergent Klebanov-Witten gauge theory being a product - $SU(N) \times SU(N)$ instead of the canonical $SU(N)$. Finally, we complete this study with a systematic and detailed construction of the spectrum of small fluctuations about the giant graviton configuration. Curiously, we find that the fluctuation spectrum depends on the size of the giant. The similarity of the operator structures in the Klebanov-Witten and ABJM theories leads us to believe that the D4-brane giant graviton in type IIA string theory on $AdS_{4}\times \mathbb{CP}^{3}$ factorizes into two $\mathbb{CP}^{2}$ dibaryons in a qualitatively similar way.}

 \preprint{CGG-130110}

\newcommand{\p}{\partial}
\begin{document}

\maketitle


\section{Introduction} \label{section - introduction}

The AdS/CFT correspondence, in its original form \cite{Maldacena}, is a conjectured duality between an $\mathcal{N}=4$ Super Yang-Mills (SYM) theory, with an $SU(N)$ gauge group, in 3+1 dimensions and type IIB string theory on $AdS_{5}\times S^{5}$. This is a strong/weak coupling duality - the t'Hooft coupling of the gauge theory is matched to the radius of the
anti-de Sitter and 5-sphere spaces\footnote{We make use of units in which $\alpha'\equiv 1$.}:
\begin{equation}
  \lambda \equiv g_{YM}^{2}N = R^{4}.
\end{equation}
In recent years, substantial progress has been made in developing the dictionary between the two sides of the correspondence. For example, the anomalous dimensions of $\tfrac{1}{2}$-BPS  (and a small class of nearly $\tfrac{1}{2}$-BPS) operators, which are protected from strong coupling corrections by supersymmetry, can now be matched quite explicitly to the energies of certain semi-classical string states \cite{GKP and Witten,BMN}.  In fact, a convincing body of evidence has begun to emerge\footnote{For a recent concise and introductory review, see \cite{JM-RdMK}.} that, organizing these $\tfrac{1}{2}$-BPS operators according to their $\mathcal{R}$-charge encodes a gauge theory description of gravitons \cite{RdMK1}, strings \cite{BMN}, D-branes \cite{CJR, Berenstein, RdMK-et-al} and even whole new geometries \cite{RdMK1,LLM geometries}.

Our interest here will be in a particular type of D-brane called a {\it giant graviton}. Giant gravitons are (classically) stable D-brane configurations in the string theory that wrap some contractible cycle in the 10-dimensional geometry. They are stabilized by their coupling to the 5-form field strength\footnote{Here we have in mind D3-brane giant gravitons.  D2-brane and D4-brane giants will couple to 4-form and 6-form field strengths respectively.} which produces a Lorentz-like force that balances the brane tension. In the gauge theory these giant graviton states are dual to operators with $\mathcal{R}$-charge of $O(N)$, constructed as a {\it Schur polynomial} function of the complex scalars $X$, $Y$ and $Z$ that constitute the Higgs sector of the SYM theory \cite{BBNS,CJR,Berenstein}. For example, restricting to a single-matrix ($Z$) model
\begin{equation}
\chi_{R}(Z) = \frac{1}{n!}\sum_{\sigma \epsilon S_{n}}\chi_{R}(\sigma)~Z^{i_{1}}_{i_{\sigma(1)}}~Z^{i_{2}}_{i_{\sigma(2)}}\ldots ~ Z^{i_{n}}_{i_{\sigma(n)}}.
\end{equation}
Here $\sigma$ is an element of the permutation group $S_{n}$, with character $\chi_{R}(\sigma)$, in the representation $R$. In other words\footnote{Quite a few other words actually. See, for example, the series of articles \cite{RdMK-et-al} for a step-by-step development of this remarkable technology.}, by the Schur-Weyl duality, the operators dual to giant gravitons are labelled by Young diagrams that encode the representations of $S_{n}$. Not only do these gauge invariant Schur polynomials diagonalize the free two-point functions and provide a good basis for the ${1\over 2}$-BPS sector of ${\cal N}=4$ super Yang-Mills theory, but they also realize quite explicitly some of the characteristic properties of the dual D-branes \cite{Balasubramanian:2004nb,DJM,Berenstein-et-al,RdMK-et-al,HM,IS-RdMK,Pirrone}.

For example, depending on the 3-cycle the D3-brane wraps, giant gravitons in $AdS_{5}\times S^{5}$ come in two flavours: AdS and sphere giants \cite{GMT}. These correspond to Schur polynomials in the totally symmetric and totally anti-symmetric representations of the permutation group respectively.  The latter can be written equivalently as the subdeterminant
\begin{equation}
\mathcal{O}_{n}(Z) = \frac{1}{n!} ~ \epsilon_{\alpha_{1}\ldots\alpha_{n}\alpha_{n+1}\ldots\alpha_{N}} ~ \epsilon^{\beta_{1}\ldots\beta_{n}\alpha_{n+1}\ldots\alpha_{N}} ~ Z^{\alpha_{1}}_{\beta_{1}} ~ \ldots ~ Z^{\alpha_{n}}_{\beta_{n}}.
\end{equation}
Notice the upper bound $n \leq N$ on the $\mathcal{R}$-charge of this subdeterminant operator - the size of the sphere giant (which is proportional to the square root of its angular momentum  $J$) is limited by the radius of the 5-sphere. This provides a realization of the so-called stringy exclusion principle \cite{MS}. Moreover, the combinatorics of attaching additional `words' (built out of the $X,Y$ and $Z$'s) to these operators (and to the Schur polynomials in general) encodes the Gauss law constraint satisfied by the spherical D-brane \cite{Balasubramanian:2004nb}. In this sense, the compact topology of the D-brane can be seen
as an {\it emergent} property of the gauge theory. But there's more; in a remarkable series of articles \cite{RdMK1,RdMK-et-al} it was argued, through a systematic study of open strings attached to giant gravitons and the associated dual restricted Schur polynomials, that even {\it local geometric} structure is encoded in the gauge theory. Of course, it could be suggested that we're reading too much from the $1\over 2$-BPS sector. This may well be the case, but to do better we would have to go beyond the single matrix model by understanding the full multi-matrix problem. In light of the difficulty of this task, another more modest route to testing the duality would be to accumulate more circumstantial evidence. To this end, the bigger the laboratory in which we can conduct these tests, the better.

Fortunately, following the recent construction by Bagger \& Lambert and, independently, Gustavsson (BLG) of the non-abelian worldvolume gauge theory of multiple M2-branes \cite{BLG}, a new example of an AdS/CFT correspondence was proposed by Aharony, Bergman, Jafferis \& Maldacena (ABJM) \cite{ABJM}.  In the weak-coupling limit, this relates type IIA string theory on $AdS_{4}\times \mathbb{CP}^{3}$ to an $\mathcal{N}=6$ Super Chern-Simons (SCS)-matter theory, with a $U(N)\times U(N)$ gauge group, in 2+1 dimensions.  This is again a strong/weak coupling duality, with a natural t'Hooft coupling
\begin{equation}
\lambda \equiv \frac{N}{k} = \frac{R^{4}}{2\pi^{2}},
\end{equation}
where $k$ is the Chern-Simons level number.  The matter part of the ABJM theory consists of two chiral superfields $A_1$ and $A_2$ transforming in the bifundamental $(\mathbf{N},\overline{\mathbf{N}})$ representation,
two chiral superfields $B_{1}$ and $B_{2}$ transforming in the anti-bifundamental
$(\overline{\mathbf{N}},\mathbf{N})$ representation and a superpotential
\begin{eqnarray}
  W=\frac{4\pi}{k}~\mathrm{tr}\left(A_{1}B_{1}A_{2}B_{2}-A_{1}B_{2}A_{2}B_{1}\right),
  \nonumber
\end{eqnarray}
which exhibits an explicit $SU(2) \times SU(2)$ symmetry acting separately on the A's and B's. Actually, the theory also has another symmetry which does not commute with the $SU(2) \times SU(2)$; this is an $SU(2)_R$ symmetry, under which the bosonic fields $\left(A_{1}, \bar{B}_{1}\right)$ and $\left(A_{2}, \bar{B}_{2}\right)$ transform as doublets. These two symmetries together generate an $SU(4)$, under which the four bosonic components of the $SU(4)$-valued $Y^A = (A_1, A_2, B^{\dag}_{\dot{1}},B^{\dag}_{\dot{2}})$ transform in the {\bf 4} of $SU(4)$ while those of $Y^{\dag}_A = (A^{\dag}_1, A^{\dag}_2, B_{\dot{1}},B_{\dot{2}})$ transform in the $\bar{{\bf 4}\,}$. The closed string sector of this theory has received considerable interest in the past few months. In particular, detailed studies of semiclassical strings, the integrability of the string sigma model, the near-flat and pp-wave limits as well as the giant magnon sector of the background have all been successfully carried out (see for example
\cite{Semiclassical,Integrability,Near-flat,pp-wave,Giant-Magnon} and references therein). The open string sector, on the other hand, remains largely unexplored. Since, however, it encodes all the information about D-branes and their dynamics, understanding this sector is vital. What we {\it do} know of open strings on this background reveals a remarkably rich structure
\cite{BT,BB,TN,Berenstein-Park,HMPS}, including new spinning M2-brane solutions, possible new supersymmetric black rings in $AdS_{4}$ and even giant tori.

This last configuration is particularly interesting. Not only should the operator dual to this toroidal configuration encode Gauss' law for a compact object, but it must also necessarily  reproduce its nonzero genus. This would be a decidedly nontrivial test of the idea that topology is an emergent property in the gauge theory. Initial steps towards developing this idea were taken in \cite{Berenstein-Park, Kim-Kim-Lee}.

Then there is the giant graviton `dual' to the spherical D2-brane giant: a D4-brane blown up on some trivial 4-cycle in $\mathbb{CP}^{3}$. In principle, it is clear how to construct such a configuration. In practice, the nontrivial geometry of $\mathbb{CP}^{3}$ makes this a bit more challenging. Curiously, in this case, armed with the technology developed for giant gravitons in $AdS_{5}\times S^{5}$, the construction of giant graviton operators in the gauge theory is perhaps more transparent than in the string theory. To construct a $\tfrac{1}{2}$-BPS operator dual to a giant graviton in the type IIA string theory, we simply replace $Z$ by the combination $A_{1}B_{1}$, say, in the Schur polynomial $\chi_{R}$.  This has at least one interesting consequence when $R$ is the completely antisymmetric representation: at maximal size $n=N$, the subdeterminant operator     $\mathcal{O}_{n}(A_{1}B_{1})$ dual to a giant graviton extended in the compact $\mathbb{CP}^{3}$ space factorizes into the product of two determinant operators:
\begin{eqnarray}
\nonumber & \mathcal{O}_{N}(A_{1}B_{1}) & = \frac{1}{N!} ~ \epsilon_{\alpha_{1}\ldots\alpha_{N}}(A_{1})^{\alpha_{1}}_{~\gamma_{1}}\ldots (A_{1})^{\alpha_{N}}_{~\gamma_{N}}
~ \epsilon^{\beta_{1}\ldots\beta_{N}}(B_{1})_{\beta_{1}}^{~\gamma_{1}}\ldots (B_{1})_{\beta_{N}}^{~\gamma_{N}} \\
&& = (\det A_{1})(\det B_{1}).
\end{eqnarray}
These {\it dibaryon} operators $\det{A_{1}}$ and $\det{B_{1}}$ are dual to D4-branes wrapped on different $\mathbb{CP}^{2}$ subspaces in $\mathbb{CP}^{3}$ \cite{BT,BB,MP}. Unlike giant gravitons though, each of the dibaryons is {\it topologically stable}. Among other things, we would like to know how this transition from the dynamically stable giant graviton to the topologically stable dibaryons happens.

As a `warmup' exercise\footnote{In much the same way that climbing Mount Everest might be considered a warmup for K2.}, we will study another situation where giant gravitons are expected to be intimately related to dibaryons: type IIB string theory on $AdS_{5}\times T^{1,1}$.  This string theory is dual to Klebanov-Witten theory \cite{Klebanov-Witten} - a non-renormalizable $\mathcal{N}=1$ SYM theory, with an $SU(N)\times SU(N)$ gauge group, in 3+1 dimensions, where
\begin{equation}
N = \frac{4R^{4}}{27\pi}
\end{equation}
corresponds to the number of units of RR-flux on the Einstein-Sasaki space $T^{1,1}$.  Klebanov-Witten theory again contains two types of scalar fields, $A_{i}$ and $B_{i}$, which transform in the $(\mathbf{N},\bar{\mathbf{N}})$ and $(\bar{\mathbf{N}},\mathbf{N})$ representations respectively.  There is an $SU(2)\times SU(2) \times U(1)_{\mathcal{R}}$ symmetry group, with one $SU(2)$ acting on the $A_{i}$'s and the other on the $B_{i}$'s.  The operator structure is the same as that of ABJM theory.

A D3-brane giant graviton extended in $T^{1,1}$ is dual to the subdeterminant operator $\mathcal{O}_{n}(A_{1}B_{1})$, with\footnote{The $\mathcal{R}$-charge and conformal dimension of both types of scalar fields, $A_{i}$ and $B_{i}$, are $\tfrac{1}{2}$ and $\Delta = \tfrac{3}{4}$ respectively.  This can be obtained by noticing that the superpotential in Klebanov-Witten theory takes the form $W = \tfrac{\lambda}{2}~\epsilon^{ij}\epsilon^{kl}~\textrm{tr}(A_{i}B_{j}A_{k}B_{l})$ and must have $\mathcal{R}$-charge 2.} $\mathcal{R}$-charge $n$ and conformal dimension $\Delta = \tfrac{3}{2}n$.  Based on the elegant holomorphic curve construction of \cite{Mikhailov}, we show how to construct giant gravitons in $T^{1,1}$ and realize the factorization of the maximal giant graviton into the two dibaryons of \cite{BHK}.

This paper is organized as follows: Section \ref{section - background} contains a short discussion of type IIB string theory on $AdS_{5}\times T^{1,1}$. The giant graviton is explicitly constructed in section \ref{section - gg}, together with its D3-brane action and energy.  Small fluctuations about the giant are considered in section \ref{section - fluct}. Section \ref{section - excitations} contains an analysis of short open strings attached to the giant graviton in two distinct pp-wave backgrounds associated with null geodesics on its worldvolume.  Concluding remarks are presented in section \ref{section - conclusion}

\newpage


\section{Type IIB string theory on $AdS_{5}\times T^{1,1}$} \label{section - background}

The 4-dimensional cone $\mathcal{C}$, in which the manifold $T^{1,1}$ is embedded, is described \cite{Candelas} by the complex coordinates $z^{A}$ satisfying $z^{1}z^{2} = z^{3}z^{4}$. These may be parameterized as follows:
\begin{eqnarray}
& z^{1} = r_{\mathcal{C}}^{\frac{3}{2}} ~ \sin{\tfrac{\theta_{1}}{2}}\sin{\frac{\theta_{2}}{2}}~e^{\frac{1}{2}i\left(\psi-\phi_{1}-\phi_{2}\right)} &
~~~~ \longrightarrow ~~ A_{1}B_{1} \label{par-z1} \\
& z^{2} = r_{\mathcal{C}}^{\frac{3}{2}} ~ \cos{\tfrac{\theta_{1}}{2}}\cos{\frac{\theta_{2}}{2}}~e^{\frac{1}{2}i\left(\psi+\phi_{1}+\phi_{2}\right)} &
~~~~ \longrightarrow ~~ A_{2}B_{2} \label{par-z2} \\
& z^{3} = r_{\mathcal{C}}^{\frac{3}{2}} ~ \cos{\tfrac{\theta_{1}}{2}}\sin{\frac{\theta_{2}}{2}}~e^{\frac{1}{2}i\left(\psi+\phi_{1}-\phi_{2}\right)} &
~~~~ \longrightarrow ~~ A_{2}B_{1} \label{par-z3} \\
& z^{4} = r_{\mathcal{C}}^{\frac{3}{2}} ~ \sin{\tfrac{\theta_{1}}{2}}\cos{\frac{\theta_{2}}{2}}~e^{\frac{1}{2}i\left(\psi-\phi_{1}+\phi_{2}\right)} &
~~~~ \longrightarrow ~~ A_{1}B_{2}. \label{par-z4}
\end{eqnarray}
We obtain the base manifold $T^{1,1}$ by setting $r_{\mathcal{C}} \equiv 1$.  The equation of the cone $\mathcal{C}$ suggests that we associate the complex directions $z^{A}$ with combinations of the scalar fields $A_{i}$ and $B_{i}$ in Klebanov-Witten theory as indicated above.

Type IIB string theory on $AdS_{5}\times T^{1,1}$ has the background metric
\begin{equation}
R^{-2}ds^{2} = ds_{AdS_{5}}^{2} + ds_{T^{1,1}}^{2},
\end{equation}
where
\begin{eqnarray}
& \!\! ds_{AdS_{5}}^{2} & = -\left(1+r^{2}\right) dt^{2} + \frac{dr^{2}}{\left(1+r^{2}\right)} + r^{2}d\Omega_{3}^{2} \\
\nonumber & \!\! ds_{T^{1,1}}^{2} & = \frac{1}{9}\left[d\psi + \cos{\theta_{1}}d\phi_{1} + \cos{\theta_{2}}d\phi_{2}\right]^{2}
+ \frac{1}{6}\left(d\theta_{1}^{2} + \sin^{2}{\theta_{1}}d\phi_{1}^{2}\right)
+ \frac{1}{6}\left(d\theta_{2}^{2} + \sin^{2}{\theta_{2}}d\phi_{2}^{2}\right), \\
\end{eqnarray}
with $R$ the radius of the $AdS_{5}$ and $T^{1,1}$ spaces.  Here $\psi ~ \epsilon ~ [0,4\pi)$, $\phi_{i} ~ \epsilon ~ [0,2\pi)$ and
$\theta_{i} ~ \epsilon ~ [0,\pi)$.  The two 2-spheres $(\theta_{i},\phi_{i})$ are non-trivially fibred over a $U(1)$ direction described by the angular coordinate $\psi$.

There is no dilaton field.  The 5-form field strength is $F_{5} = \mathcal{F} + \ast\mathcal{F}$, where
\begin{equation}
\mathcal{F} \equiv 4R^{4}~\textrm{vol}\left(T^{1,1}\right)
= \frac{R^{4}}{27}\sin{\theta_{1}}\sin{\theta_{2}}~d\theta_{1} \wedge d\theta_{2} \wedge d\psi \wedge d\phi_{1} \wedge d\phi_{2}.\\
\end{equation}


\section{Giant gravitons on $\mathbb{R}\times T^{1,1}$} \label{section - gg}

\subsection{Giant graviton ansatz via a holomorphic curve on the cone} \label{section - gg - curve}

We shall use the construction of \cite{Mikhailov} in terms of holomorphic curves to write down an explicit ansatz for the giant graviton, which is dual to the subdeterminant operator $\mathcal{O}_{n}(A_{1}B_{1})$, extended and moving on $\mathbb{R}\times T^{1,1}$.

Let us choose the holomorphic curve on the cone $\mathcal{C}$ to be
\begin{equation}
F(z^{A}) = z^{1} = \rho,
\end{equation}
where $\rho$ is a real constant. The surface of the giant graviton is the intersection of the holomorphic curve with the base manifold $T^{1,1}$.  For this intersection to be non-empty, $\rho$ must be confined to the unit interval, so that we may define $\rho \equiv \sqrt{1-\alpha^{2}}$, with $\alpha ~ \epsilon ~ [0,1]$. Here $\alpha$ may be thought of as the `size' of the giant graviton.

To introduce motion, take $z^{A} \rightarrow z^{A}e^{-i\varphi(t)}$ in the holomorphic function $F(z^{A})$, with $\varphi(t)$ an overall time-dependent phase\footnote{The preferred direction of \cite{Mikhailov}, induced by the embedding of $T^{1,1}$ into the cone $\mathcal{C}$, is along the fibre $\psi$.  This is independent of which holomorphic function is chosen to construct a particular giant graviton and should not be confused with the direction of motion, which is the component of the preferred direction perpendicular to the giant graviton's surface.}. Hence
\begin{equation}
F(z^{A}e^{-i\varphi(t)}) \equiv z^{1}e^{-i\varphi(t)} = \sqrt{1-\alpha^{2}} ~~~~~~ \Rightarrow ~~~ z^{1} = \sqrt{1-\alpha^{2}}~ e^{i\varphi(t)},
\end{equation}
where we hold the other \emph{independent} coordinates fixed. There is a subtlety involved, however, in choosing which two complex coordinates on the cone, aside from $z^{1}$, to consider as independent.  These should correspond to exactly those angular directions along which the giant graviton does \emph{not} rotate.  Since the dual operators are constructed out of equal numbers of $A_{1}$'s and $B_{1}$'s, we see that $z^{1}$, $z^{2}$ and $z^{3}/z^{4}$ (or $z^{4}/z^{3}$) are the correct independent coordinates to use.  Therefore, we shall rotate along the $\tfrac{1}{2}(\psi-\phi_{1}-\phi_{2})$ direction, while holding $\tfrac{1}{2}(\psi +\phi_{1} + \phi_{2})$ and $\phi_{1} - \phi_{2}$ fixed.

We shall now define
\begin{eqnarray}
&& \chi_{1} \equiv \tfrac{1}{3}\left(\psi - \phi_{1} - \phi_{2}\right), \label{chi1} \\
&& \chi_{2} \equiv \tfrac{1}{3}\left(\psi + 3\phi_{1} -\phi_{2}\right)
= \tfrac{2}{3}\left[\tfrac{1}{2}\left(\psi +\phi_{1} + \phi_{2}\right) + \left(\phi_{1} - \phi_{2}\right)\right], \label{chi2} \\
&& \chi_{3} \equiv \tfrac{1}{3}\left(\psi - \phi_{1} + 3\phi_{2}\right)
= \tfrac{2}{3}\left[\tfrac{1}{2}\left(\psi +\phi_{1} + \phi_{2}\right) - \left(\phi_{1} - \phi_{2}\right)\right]. \label{chi3}
\end{eqnarray}
Note that $\chi_{2}$ and $\chi_{3}$ are combinations of the phases of our independent coordinates $z^{2}$ and $z^{3}/z^{4}$. The complex coordinates $z^{A}$, confined to the base manifold $T^{1,1}$, can be written as
\begin{eqnarray} \label{par-z-chi}
\nonumber && z^{1} = \sin{\tfrac{\theta_{1}}{2}}\sin{\tfrac{\theta_{2}}{2}}~e^{\frac{3}{2}i\chi_{1}}, \hspace{0.85cm}  ~~~~~~~~~~
z^{2} = \cos{\tfrac{\theta_{1}}{2}}\cos{\tfrac{\theta_{2}}{2}}~e^{\frac{3}{4}i(\chi_{2}+\chi_{3})}, \\
&& z^{3} = \cos{\tfrac{\theta_{1}}{2}}\sin{\tfrac{\theta_{2}}{2}}~e^{\frac{3}{4}i(\chi_{1}+\chi_{2})}, ~~~~~~~~~~
z^{4} = \sin{\tfrac{\theta_{1}}{2}}\cos{\tfrac{\theta_{2}}{2}}~e^{\frac{3}{4}i(\chi_{1}+\chi_{3})}.
\end{eqnarray}
The giant graviton ansatz then translates into setting
\begin{equation} \label{def-alpha}
\sin{\tfrac{\theta_{1}}{2}}\sin{\tfrac{\theta_{2}}{2}} = \sqrt{1-\alpha^{2}},
\end{equation}
and considering the angular direction of motion $\chi_{1}(t)$.

\bigskip

\textbf{\emph{Point graviton} ($\alpha = 0$)}

When $\alpha = 0$, we obtain the point graviton.  Here $z^{1}=e^{\frac{3}{2}i\chi_{1}(t)}$ and $z^{2}=z^{3}=z^{4}=0$, which describes the motion of a point along a circle of maximum radius.

\bigskip

\textbf{\emph{Maximal giant graviton} ($\alpha = 1$)}

The maximal giant graviton is obtained by setting $\alpha = 1$.  In this case, the polar coordinates $\theta_{1}$ and $\theta_{2}$ on each of the 2-spheres decouple and we find two distinct solutions
\begin{equation}
\sin{\tfrac{\theta_{2}}{2}} = 0 ~~~~~~~~ \textrm{or} ~~~~~~~~ \sin{\tfrac{\theta_{1}}{2}} = 0,
\end{equation}
which describes the union of the two spaces $\theta_{2}=0$ or $\theta_{1}=0$.  These are the dibaryons of \cite{BHK} - two D3-branes wrapped on different 2-spheres and the $U(1)$ fibre - corresponding to the determinant operators $\det{A_{1}}$ and $\det{B_{1}}$ respectively.

\bigskip

\textbf{\emph{Submaximal giant graviton}}

We would like to understand this factorization into two dibaryons as some submaximal giant graviton configuration in the limit as $\alpha \rightarrow 1$. Key in this endeavour is our choice of worldvolume coordinates: the obvious independent angles are $\chi_{2}$ and $\chi_{3}$, but how do we choose a radial parameter describing the giant graviton worldvolume? To obtain the maximal giant as a limiting case, we cannot choose either $\theta_{1}$ or $\theta_{2}$, as this choice would eliminate half the maximal giant a priori. Let us rather consider the combination
\begin{equation} \label{def-u}
u = \cos{\tfrac{\theta_{1}}{2}}\cos{\tfrac{\theta_{2}}{2}},
\end{equation}
which is the magnitude of the complex coordinate $z^{2}$.  Using the relation \ref{def-alpha} between $\theta_{1}$ and $\theta_{2}$ on the giant graviton worldvolume, we can rewrite
\begin{equation}
u(\theta_{i}) = \cot{\tfrac{\theta_{i}}{2}}\sqrt{\alpha^{2} - \cos^{2}{\tfrac{\theta_{i}}{2}}}.
\end{equation}
Note that $\theta_{i}$ is only defined on the interval $[2\arccos{\alpha},\pi]$.  The function $u(\theta_{i})$ vanishes at both ends of this interval and attains a maximum value $u_{\textrm{max}} = 1-\sqrt{1-\alpha^{2}}$ at the polar angle $\theta_{i,\textrm{max}} = 2\arcsin{(1-\alpha^{2})^{1/4}}$.

\begin{figure}[htb!]
\begin{center}
\includegraphics[width = 15.0cm, height = 4.62cm]{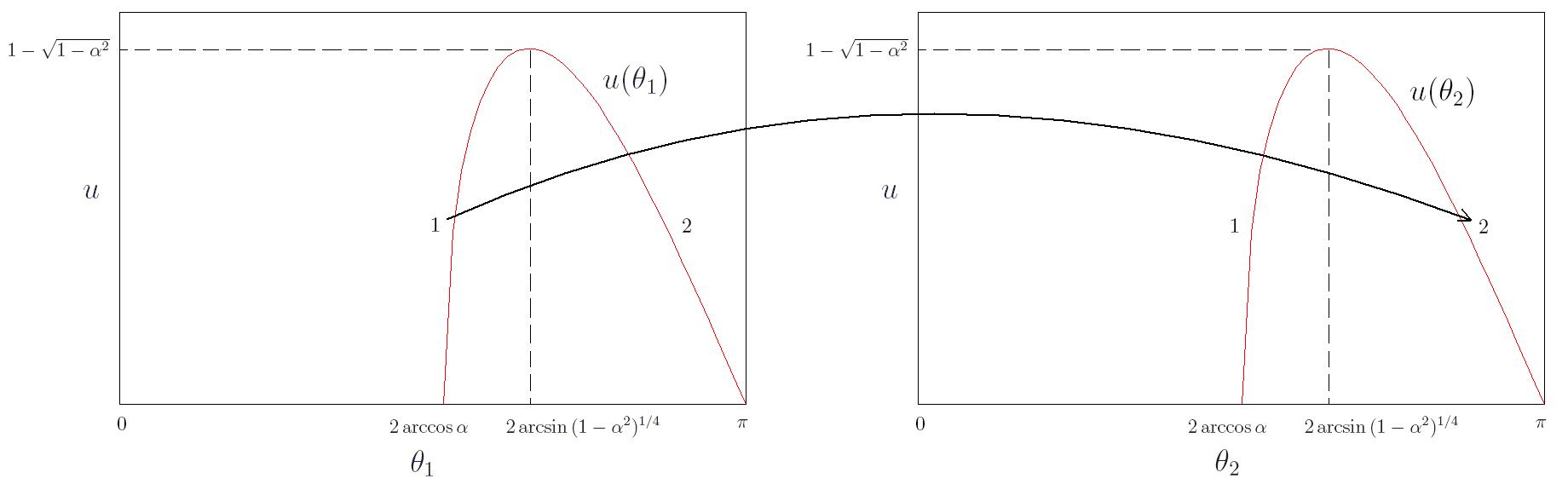} 
\caption{The radial coordinate $u$, confined to the worldvolume of the giant graviton on which $\alpha$ is constant, plotted as a function of the polar angles $\theta_{1}$ and $\theta_{2}$ respectively.  The mapping between the $u(\theta_{1})$ and $u(\theta_{2})$ diagrams is shown.} \label{u mapping}
\end{center}
\end{figure}

Now, we observe that the worldvolume of the giant graviton is a double-covering of $u$.  The $\theta_{1}$ interval naturally splits into two pieces $[2\arccos{\alpha}, 2\arcsin{(1-\alpha^{2})^{1/4}}]$ and $[2\arcsin{(1-\alpha^{2})^{1/4}},\pi]$, which, since the $u(\theta_{1})$ maximum occurs when $\theta_{1}=\theta_{2}$, simply correspond to $\theta_{1}\leq\theta_{2}$ and $\theta_{1}\geq\theta_{2}$. Leaving the second interval in terms of $\theta_{1}$, one could choose to describe the first region in terms of $\theta_{2}$, mapping\footnote{Note that there is a change in orientation under this map.} it onto the second interval in the analogous $u(\theta_{2})$ diagram (see figure \ref{u mapping}).

The D3-brane action for the submaximal giant graviton will therefore consist of two identical parts, involving integrals over $\theta_{1}$ and $\theta_{2}$ respectively, which run from $2\arcsin{(1-\alpha^{2})^{1/4}}$ to $\pi$ (although we shall find it more convenient to simply double the integral over $u$ from 0 to $1-\sqrt{1-\alpha^{2}}$).  Note that this action still describes a single D3-brane extended on both 2-spheres.  In the limit $\alpha\rightarrow 1$, each of the second $\theta_{i}$ regions covers an entire 2-sphere, whilest the first completely vanishes.  In this way, we recover both halves of the maximal giant graviton.

\begin{figure}[htb!]
\begin{center}
\includegraphics[width = 16cm, height = 3cm]{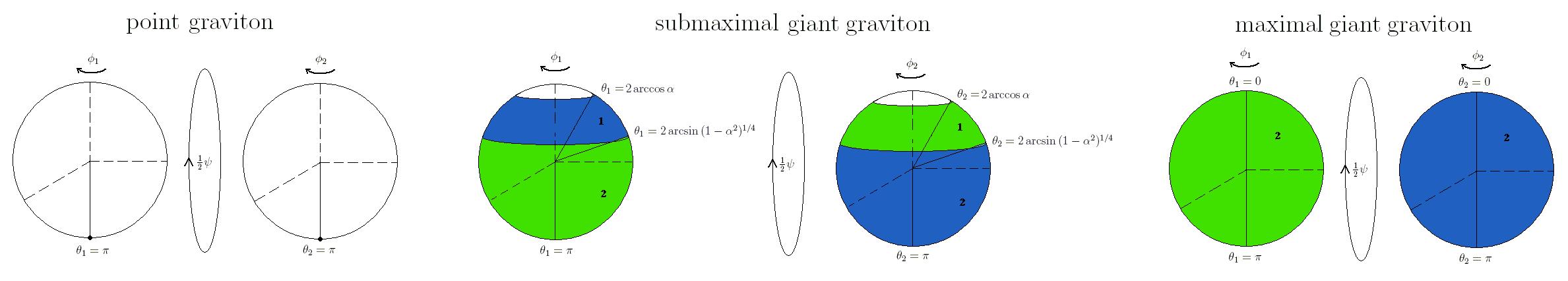}   
\caption{Pictorial representation of the expansion of a point graviton, via a submaximal giant graviton intermediate state, into the maximal giant graviton in $\mathbb{R}\times T^{1,1}$. Regions identically shaded (either blue or green) are mapped onto each other by the constraint $\sin{\tfrac{\theta_{1}}{2}}\sin{\tfrac{\theta_{2}}{2}} = \sqrt{1-\alpha^{2}}$, which describes the worldvolume of the giant graviton. The factorization of the maximal giant into two dibaryons is clearly visible.} \label{submax giant}
\end{center}
\end{figure}

This construction allows us to see the intermediate state - the submaximal giant graviton - between the point graviton and the maximal giant graviton. We observe the manner in which the maximal giant factorizes - the relation between $\theta_{1}$ and $\theta_{2}$, and the mapping between the different regions (shaded the same colour in figure \ref{submax giant}) of the two 2-spheres disappears.


\subsection{Angular and radial coordinate changes} \label{section - gg - angular and radial}

The giant graviton is a D3-brane in type IIB string theory on $AdS_{5}\times T^{1,1}$ situated at the center of the $AdS_{5}$ spacetime (moving only in time $t$). We may therefore restrict ourselves to the background $\mathbb{R}\times T^{1,1}$, which has the following metric:
\begin{equation}
R^{-2} ds^{2} = -dt^{2} + ds_{\textrm{radial}}^{2} + ds_{\textrm{angular}}^{2},
\end{equation}
with
\begin{eqnarray}
&& \!\! ds_{\textrm{radial}}^{2} = \tfrac{1}{6}\left\{d\theta_{1}^{2} + d\theta_{2}^{2}\right\} \label{ds2-rad} \\
&& \!\! ds_{\textrm{angular}}^{2} = \tfrac{1}{18}\left\{2\left(d\psi + \cos{\theta_{1}}d\phi_{1} + \cos{\theta_{2}}d\phi_{2}\right)^{2} + 3\sin^{2}{\theta_{1}}d\phi_{1}^{2} + 3\sin^{2}{\theta_{2}}d\phi_{2}^{2}\right\}, ~~~~~~~~ \label{ds2-ang}
\end{eqnarray}
describing the radial and angular parts of the metric (separated for later convenience) associated with the magnitudes and phases of our complex coordinates $z^{A}$, which parameterize $T^{1,1}$.

The giant graviton couples to the 4-form potential with corresponding $T^{1,1}$ field strength
\begin{equation}
F_{5} = \frac{R^{4}}{27}\sin{\theta_{1}}\sin{\theta_{2}} ~ d\theta_{1} \wedge d\theta_{2} \wedge d\psi \wedge d\phi_{1} \wedge d\phi_{2}.
\end{equation}

\bigskip

\textbf{\emph{Angular coordinates $\chi_{i}$}}

We shall now change to the angular coordinates $\chi_{i}$, defined in \ref{chi1} - \ref{chi3}, which most conveniently parameterize the direction of motion and angular extension of the giant graviton.  The components of the angular metric $(g_{\chi})_{ij}$ are stated explicitly in appendix \ref{appendix - conifold metric} and the determinant is given by
\begin{equation}
\det{g}_{\chi} = \left(\tfrac{3}{32}\right)^{2}\sin^{2}{\theta_{1}}\sin^{2}{\theta_{2}}.
\end{equation}
We shall also need to know the determinant of the angular metric restricted to the worldvolume coordinates $\chi_{2}$ and $\chi_{3}$:
\begin{equation}
\left(C_{\chi}\right)_{11} = 3\left(\tfrac{1}{32}\right)^{2}\left\{2\sin^{2}{\theta_{1}}\left(1+\cos{\theta_{2}}\right)^{2} + 2\sin^{2}{\theta_{2}}\left(1+\cos{\theta_{1}}\right)^{2} + 3\sin^{2}{\theta_{1}}\sin^{2}{\theta_{2}}\right\},
\end{equation}
which is also the cofactor of the element $\left(g_{\chi}\right)_{11}$.

The 5-form field strength in terms of these angular coordinates $\chi_{i}$ (and the original radial coordinates $\theta_{i}$) is
\begin{equation}
F_{5} = \frac{R^{4}}{16}\sin{\theta_{1}}\sin{\theta_{2}}
~ d\theta_{1} \wedge d\theta_{2} \wedge d\chi_{1} \wedge d\chi_{2} \wedge d\chi_{3}.
\end{equation}

\bigskip
\newpage

\textbf{\emph{Orthogonal radial coordinates $\alpha$ and $v$}}

The coordinates $\alpha$ and $u$, which are defined by
\begin{equation}
\sqrt{1-\alpha^{2}} \equiv \sin{\tfrac{\theta_{1}}{2}}\sin{\tfrac{\theta_{2}}{2}} ~~~~~~~~ \textrm{and} ~~~~~~~~
u \equiv \cos{\tfrac{\theta_{1}}{2}}\cos{\tfrac{\theta_{2}}{2}},
\end{equation}
are well suited to describe the size and radial extension of the giant graviton, but turn out to be non-orthogonal.  We shall hence rather choose to describe the radial space in terms of the orthogonal radial coordinates $\alpha$ and $v$, with the latter defined as follows:
\begin{equation} \label{def-v}
v \equiv \frac{2u}{\alpha^{2} + u^{2}},
\end{equation}
in terms of which the radial metric can be written as
\begin{equation}
ds_{\textrm{radial}}^{2} = g_{\alpha}d\alpha^{2} + g_{v}dv^{2},
\end{equation}
with
\begin{equation} \label{g-alpha and g-v}
g_{\alpha} = \frac{\alpha^{2}v^{2}}{3\left(1-\alpha^{2}\right)\sqrt{1-\alpha^{2}v^{2}}\left(1-\sqrt{1-\alpha^{2}v^{2}}\right)}
~~~~ \textrm{and} ~~~~
g_{v} = \frac{\left(1-\sqrt{1-\alpha^{2}v^{2}}\right)}{3v^{2}\left(1-v^{2}\right)\sqrt{1-\alpha^{2}v^{2}}}.
\end{equation}

We can rewrite the determinant of the angular metric in terms of the orthogonal radial coordinates $\alpha$ and $v$ as follows:
\begin{equation} \label{det-g}
\det{g_{\chi}} = \frac{9}{64}\left(1-\alpha^{2}\right)\frac{1}{v^{2}}\left(1-\sqrt{1-\alpha^{2}v^{2}}\right)^{2},
\end{equation}
and the cofactor associated with $\left(g_{\chi}\right)_{11}$ becomes
\begin{equation} \label{C11}
\left(C_{\chi}\right)_{11} = \frac{3}{64}\frac{1}{v^{2}}\left(1-\sqrt{1-\alpha^{2}v^{2}}\right)^{2}
\left\{\frac{4}{v^{2}}\sqrt{1-\alpha^{2}v^{2}}\left(1-\sqrt{1-\alpha^{2}v^{2}}\right) + 3\left(1-\alpha^{2}\right)\right\}.
\end{equation}

The 5-form field strength is
\begin{equation}
F_{5} = \mp\frac{R^{4}}{2}\frac{\alpha\left(1-\sqrt{1-\alpha^{2}v^{2}}\right)}{v\sqrt{1-v^{2}}\sqrt{1-\alpha^{2}v^{2}}} ~
d\alpha \wedge dv \wedge d\chi_{1} \wedge d\chi_{2} \wedge d\chi_{3},
\end{equation}
and the associated 4-form potential is given by
\begin{equation} \label{C4}
C_{4} = \mp\frac{R^{4}}{4}\frac{\left(1-\sqrt{1-\alpha^{2}v^{2}}\right)^{2}}{v^{3}\sqrt{1-v^{2}}} ~
dv \wedge d\chi_{1} \wedge d\chi_{2} \wedge d\chi_{3}.
\end{equation}
where the $\mp$ distinguishes between the intervals $\theta_{1} \geq \theta_{2}$ and $\theta_{1} \leq \theta_{2}$. Notice that we have chosen a gauge in which $C_{4}$ is non-singular at $\alpha = 0$.


\subsection{D3-brane action} \label{section - gg - action}

As a dynamical object in type IIB string theory, our giant graviton is described by the D3-brane action $S = S_{DBI} + S_{WZ}$, which consists of the Dirac-Born-Infeld (DBI) and Wess-Zumino (WZ) terms
\begin{equation}
S_{DBI} = -T_{3}\int_{\Sigma} ~ d^{4}\sigma ~ \sqrt{-\det{\mathcal{P}\left[g\right]}} ~~~~~~ \textrm{and} ~~~~~~
S_{WZ} = \pm T_{3} \int_{\Sigma} ~  \mathcal{P}\left[C_{4}\right],
\end{equation}
with $T_{3}=\tfrac{1}{(2\pi)^{3}}$ the tension.  The $\pm$ simply distinguishes between a brane and an anti-brane - we shall choose to consider branes, but the results for anti-branes are identical up to a change in the direction of motion $\dot{\chi}_{1}\rightarrow -\dot{\chi}_{1}$.  Here $\mathcal{P}$ denotes the pull-back to the worldvolume $\Sigma$ of the giant graviton.  We shall choose the worldvolume coordinates to be $\sigma^{0} = t$, $\sigma^{1} = v$ (double-covered), $\sigma^{2} = \chi_{2}$ and $\sigma^{3} = \chi_{3}$.

Let us consider the DBI action.  The determinant of the pull-back of the metric satisfies
\begin{equation}
-R^{-8}\det{\mathcal{P}\left[g\right]} = g_{v}\left[\left(C_{\chi}\right)_{11} - \dot{\chi}_{1}^{2}\left(\det{g_{\chi}}\right)\right],
\end{equation}
and, using the expressions \ref{g-alpha and g-v} - \ref{C11} for the various determinants associated with the angular and radial parts of the metric, we obtain
\begin{equation}
-\det{\mathcal{P}\left[g\right]} = \frac{R^{8}}{16}\frac{\left(1-\sqrt{1-\alpha^{2}v^{2}}\right)^{4}}{v^{6}\left(1-v^{2}\right)}
\left\{1 + \frac{3}{4}\frac{\left(1-\alpha^{2}\right)v^{2}\left(1-\dot{\chi}_{1}^{2}\right)} {\sqrt{1-\alpha^{2}v^{2}}\left(1-\sqrt{1-\alpha^{2}v^{2}}\right)}\right\}.
\end{equation}

To determine the WZ term, we require the pull-back of the 4-form potential \ref{C4}, which is given by
\begin{equation}
\mathcal{P}\left[C_{4}\right] = \frac{R^{4}}{4}\frac{\left(1-\sqrt{1-\alpha^{2}v^{2}}\right)^{2}}{v^{3}\sqrt{1-v^{2}}} ~ \dot{\chi}_{1} ~
dt \wedge dv \wedge d\chi_{2} \wedge d\chi_{3}.
\end{equation}

Combining the above results, we obtain the D3-brane action, which describes a giant graviton extended and moving on $\mathbb{R}\times T^{1,1}$, as follows:
\begin{equation}
S = \frac{16\pi^{2}T_{3}R^{4}}{9}\int{dt} ~ L,
\end{equation}
with the Lagrangian
\begin{equation} \label{lag-exact}
L = \int_{0}^{1}dv ~ \frac{2\left(1-\sqrt{1-\alpha^{2}v^{2}}\right)^{2}}{v^{3}\sqrt{1-v^{2}}} ~ \left\{\dot{\chi}_{1} -\sqrt{1 + \frac{3}{4}\frac{\left(1-\alpha^{2}\right)v^{2}\left(1-\dot{\chi}_{1}^{2}\right)} {\sqrt{1-\alpha^{2}v^{2}}\left(1-\sqrt{1-\alpha^{2}v^{2}}\right)}}\right\},
\end{equation}
where we have integrated over $\chi_{2}$ and $\chi_{3}$, which vary over the range $[0,\tfrac{8\pi}{3}]$.

The conserved momentum $P_{\chi_{1}} \equiv \tfrac{\p L}{\p \dot{\chi}_{1}}$ conjugate to the angular coordinate $\chi_{1}$, which describes the direction of motion, is
\begin{equation}
P_{\chi_{1}} = \int_{0}^{1}dv ~ \frac{2\left(1-\sqrt{1-\alpha^{2}v^{2}}\right)^{2}}{v^{3}\sqrt{1-v^{2}}} \left\{ 1 + \dot{\chi}_{1} ~ \frac{\frac{3}{4}\frac{\left(1-\alpha^{2}\right)v^{2}}{\sqrt{1-\alpha^{2}v^{2}}\left(1-\sqrt{1-\alpha^{2}v^{2}}\right)}}{\sqrt{1 + \frac{3}{4}\frac{\left(1-\alpha^{2}\right)v^{2}\left(1-\dot{\chi}_{1}^{2}\right)} {\sqrt{1-\alpha^{2}v^{2}}\left(1-\sqrt{1-\alpha^{2}v^{2}}\right)}}} \right\}.
\end{equation}
The Hamiltonian $H = \dot{\chi}_{1}P_{\chi_{1}} - L$ can be explicitly written as
\begin{equation}
H = \int_{0}^{1}dv ~ \frac{2\left(1-\sqrt{1-\alpha^{2}v^{2}}\right)^{2}}{v^{3}\sqrt{1-v^{2}}}
~ \frac{\left[1 + \frac{3}{4}\frac{\left(1-\alpha^{2}\right)v^{2}}{\sqrt{1-\alpha^{2}v^{2}}\left(1-\sqrt{1-\alpha^{2}v^{2}}\right)}\right]}{\sqrt{1 + \frac{3}{4}\frac{\left(1-\alpha^{2}\right)v^{2}\left(1-\dot{\chi}_{1}^{2}\right)} {\sqrt{1-\alpha^{2}v^{2}}\left(1-\sqrt{1-\alpha^{2}v^{2}}\right)}}}.
\end{equation}
These expressions describe the momentum and energy in units of $\tfrac{16\pi^{2}T_{3}R^{4}}{9}$.  We would now like to minimize $H(\alpha,P_{\chi_{1}})$ with respect to $\alpha$ for fixed momentum $P_{\chi_{1}}$.  However, since it is not immediately obvious how to invert $P_{\chi_{1}}(\dot{\chi}_{1})$ analytically, we shall first consider certain special cases.

\bigskip

\textbf{\emph{Maximal giant graviton} ($\alpha = 1$)}

When $\alpha = 1$, it is possible to evaluate the integrals over $v$ analytically. The Lagrangian becomes $L = \dot{\chi}_{1} - 1$ and $H = P_{\chi_{1}} = 1$.  All dependence on $\dot{\chi}_{1}$ disappears from the Hamiltonian $H$ and the momentum $P_{\chi_{1}}$, which is now due entirely to the extension of the D3-brane rather than to its motion along $\chi_{1}$.

\bigskip

\textbf{\emph{Small submaximal giant graviton} ($\alpha \ll 1$)}

Let us assume that $\alpha \ll 1$, so that we are considering a small submaximal giant graviton, and expand the Lagrangian in orders of $\alpha$:
\begin{equation}
L \approx \frac{\alpha^{3}}{2}\left\{\alpha\dot{\chi}_{1} - \sqrt{\alpha^{2} + \tfrac{3}{2}\left(1-\dot{\chi}_{1}^{2}\right)}\right\}.
\end{equation}
Here we must be careful to allow for the possibility that $1-\dot{\chi}_{1}^{2}$ might be small, which is why we cannot further simplify the square root. The momentum conjugate to $\chi_{1}$ is thus
\begin{equation}
P_{\chi_{1}} \approx \frac{\alpha^{3}}{2}\left\{\alpha + \frac{\frac{3}{2}\dot{\chi}_{1}}{\sqrt{\alpha^{2} + \frac{3}{2}\left(1-\dot{\chi}_{1}^{2}\right)}}\right\},
\end{equation}
and the Hamiltonian is given by
\begin{equation}
H \approx \frac{\alpha^{3}}{2}\frac{\left(\alpha^{2}+\frac{3}{2}\right)}{\sqrt{\alpha^{2} + \frac{3}{2}\left(1-\dot{\chi}_{1}^{2}\right)}}.
\end{equation}

In this approximation of small $\alpha$, it is possible to isolate all dependence on $\dot{\chi}_{1}$ and invert the momentum. We obtain
\begin{equation}
\dot{\chi}_{1}^{2} = \frac{\left(\alpha^{2} + \frac{3}{2}\right)\left(P_{\chi_{1}}-\frac{1}{2}\alpha^{4}\right)^{2}}{\left[\frac{9}{16}\alpha^{6} + \frac{3}{2}\left(P_{\chi_{1}}-\frac{1}{2}\alpha^{4}\right)^{2}\right]}.
\end{equation}
We can write the Hamiltonian as a function of the size of the giant graviton $\alpha$ and its momentum $P_{\chi_{1}}$ as follows:
\begin{equation}
H \approx \sqrt{\tfrac{2}{3}\alpha^{2} + 1}\sqrt{\tfrac{3}{8}\alpha^{6} + \left(P_{\chi_{1}} - \tfrac{1}{2}\alpha^{4}\right)^{2}}.
\end{equation}

\begin{figure}[htb!]
\begin{center}
\includegraphics[width = 7.0cm, height = 4.375cm]{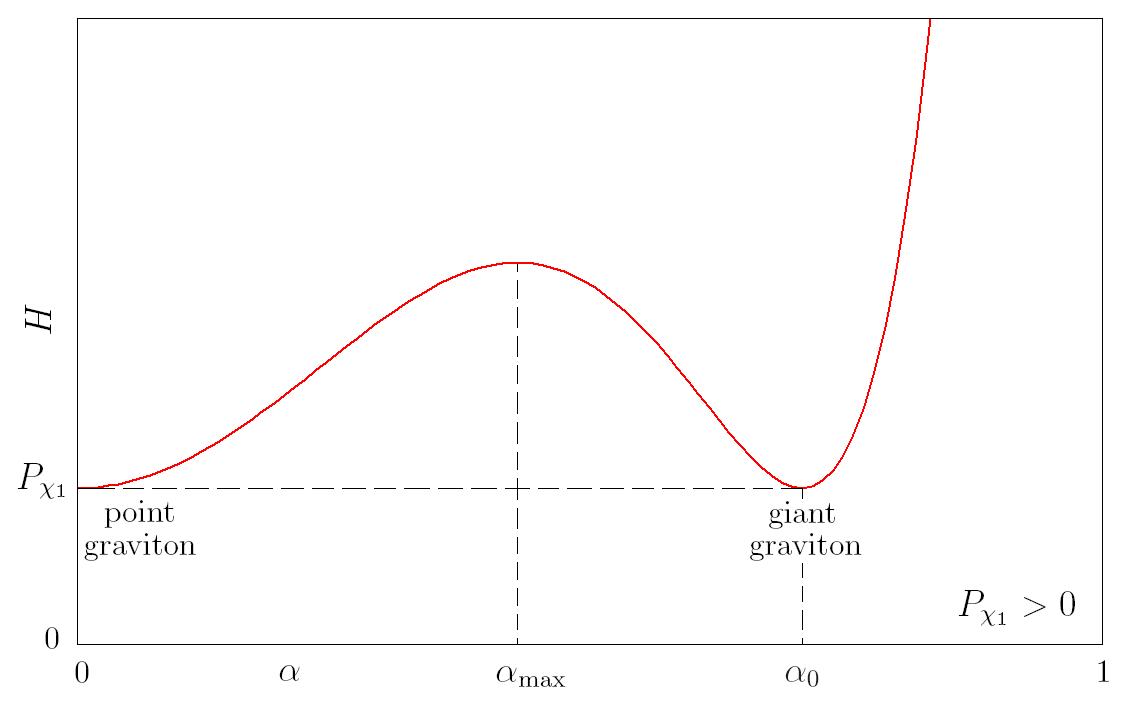}  
\includegraphics[width = 7.0cm, height = 4.375cm]{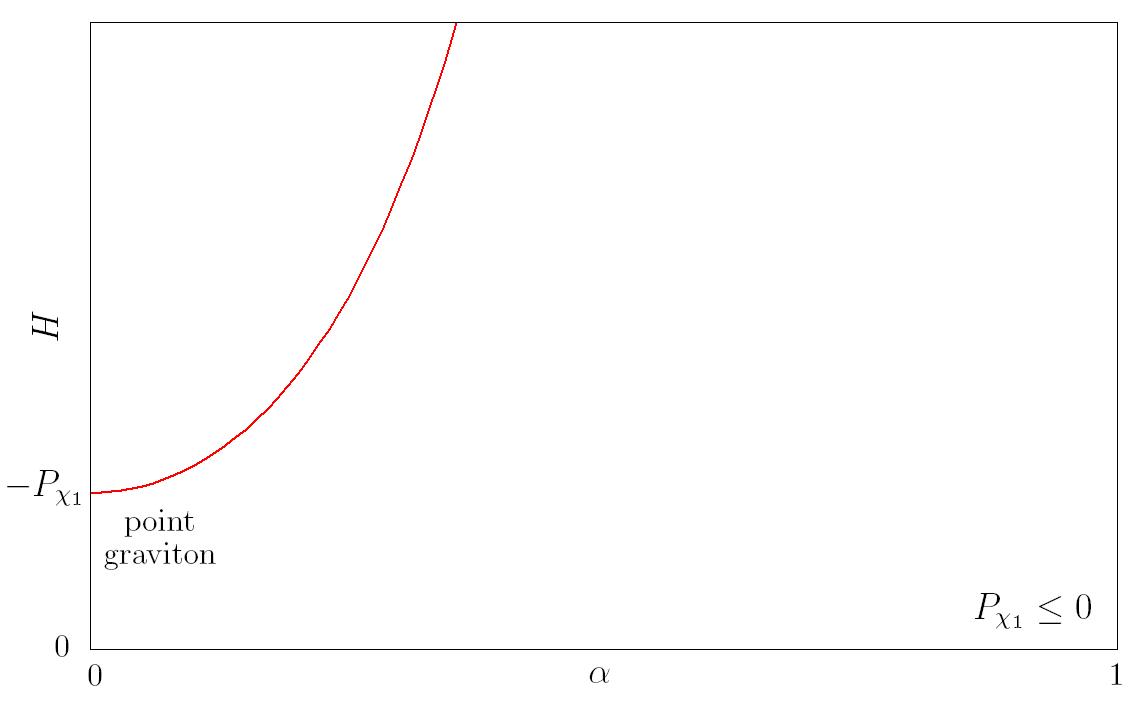}  
\caption{A generic sketch of the approximate energy $H(\alpha, P_{\chi_{1}})$, when $\alpha \ll 1$, as a function of $\alpha$ at fixed momentum $P_{\chi_{1}}$. Note that the submaximal giant graviton only exists when $P_{\chi_{1}}$ is positive and is energetically degenerate with the point graviton solution.} \label{H-approx}
\end{center}
\end{figure}

Now, it is possible to solve for the maxima and minima.  When the momentum $P_{\chi_{1}}$ is positive, this approximate energy is minimum at $\alpha = 0$ and $\alpha = \alpha_{0}$, and maximum at $\alpha = \alpha_{\textrm{max}}$ in between (see figure \ref{H-approx}). Here we define
\begin{equation}
\alpha_{0} \equiv \sqrt{\sqrt{\left(\tfrac{3}{4}\right)^{2} + 2P_{\chi_{1}}} - \tfrac{3}{4}}
~~~~~~ \textrm{and} ~~~~~~
\alpha_{\textrm{max}} \equiv \sqrt{\sqrt{\left(\tfrac{9}{20}\right)^{2} + \tfrac{2}{5}P_{\chi_{1}}} - \tfrac{9}{20}}.
\end{equation}
These minima are energetically degenerate with $H(\alpha_{\textrm{0}},P_{\chi_{1}}) \approx H(0,P_{\chi_{1}}) = P_{\chi_{1}}$.  The non-trivial minimum at $\alpha_{\textrm{0}}$ is associated with the submaximal giant graviton.  Although the expression for the Hamiltonian is approximate, the energy of the point graviton solution at $\alpha = 0$ is exact.  Furthermore, we shall later argue that the submaximal giant graviton remains degenerate with the point graviton even when its size $\alpha_{\textrm{0}}$ is large.  When the momentum $P_{\chi_{1}}$ is negative, only the trivial minimum at $\alpha = 0$, corresponding to a point graviton with energy $H(0,P_{\chi_{1}}) = -P_{\chi_{1}}$, exists.

\bigskip
\newpage

\textbf{\emph{Submaximal giant graviton}}

It turns out that, for all values of $\alpha$, the Lagrangian \ref{lag-exact} vanishes when $\dot{\chi}_{1} = 1$.  This implies that $H =  P_{\chi_{1}}$, which remains a minimum\footnote{This may be deduced by expanding $H-P_{\chi_{1}}$ in the vicinity of $\dot{\chi}_{1} = 1$ and noticing that it is always non-negative.} and corresponds to the submaximal giant graviton solution, described in the previous section when $\alpha \ll 1$.  The size of the giant $\alpha_{\textrm{0}}$ is then determined by the momentum $P_{\chi_{1}}$ via the following relation:
\begin{equation}
P_{\chi_{1}}\left(\alpha_{0}\right) = I_{1}\left(\alpha_{0}^{2}\right) + \frac{3}{4}\left(1-\alpha_{0}^{2}\right)I_{2}\left(\alpha_{0}^{2}\right),
\end{equation}
where
\begin{eqnarray}
&& I_{1}\left(\alpha_{0}^{2}\right) \equiv \int_{0}^{1}dv ~ \frac{2\left(1-\sqrt{1-\alpha_{0}^{2}v^{2}}\right)^{2}}{v^{3}\sqrt{1-v^{2}}}
= \left(1-\alpha_{0}^{2}\right)\ln{\left(1-\alpha_{0}^{2}\right)} + \alpha_{0}^{2} \\
&& I_{2}\left(\alpha_{0}^{2}\right) \equiv \int_{0}^{1}dv ~ \frac{2\left(1-\sqrt{1-\alpha_{0}^{2}v^{2}}\right)}{v\sqrt{1-v^{2}}\sqrt{1-\alpha_{0}^{2}v^{2}}} = \frac{\p I_{1}\left(\alpha_{0}^{2}\right)}{\p \left(\alpha_{0}^{2}\right)} = -\ln{\left(1-\alpha_{0}^{2}\right)}.
\end{eqnarray}
Simplifying, the exact expression for the energy and momentum of a submaximal giant graviton with size $\alpha_{0}$ is
\begin{equation}
\label{radial dependence}
H\left(\alpha_{0}\right) = P_{\chi_{1}}\left(\alpha_{0}\right) = 1 - \left(1-\alpha_{0}^{2}\right)\left[1 - \frac{1}{4}\ln{\left(1-\alpha_{0}^{2}\right)}\right],
\end{equation}
which is shown in figure \ref{H-giant} below.
\begin{figure}[htb!]
\begin{center}
\includegraphics[width = 7.5cm, height = 5.0cm]{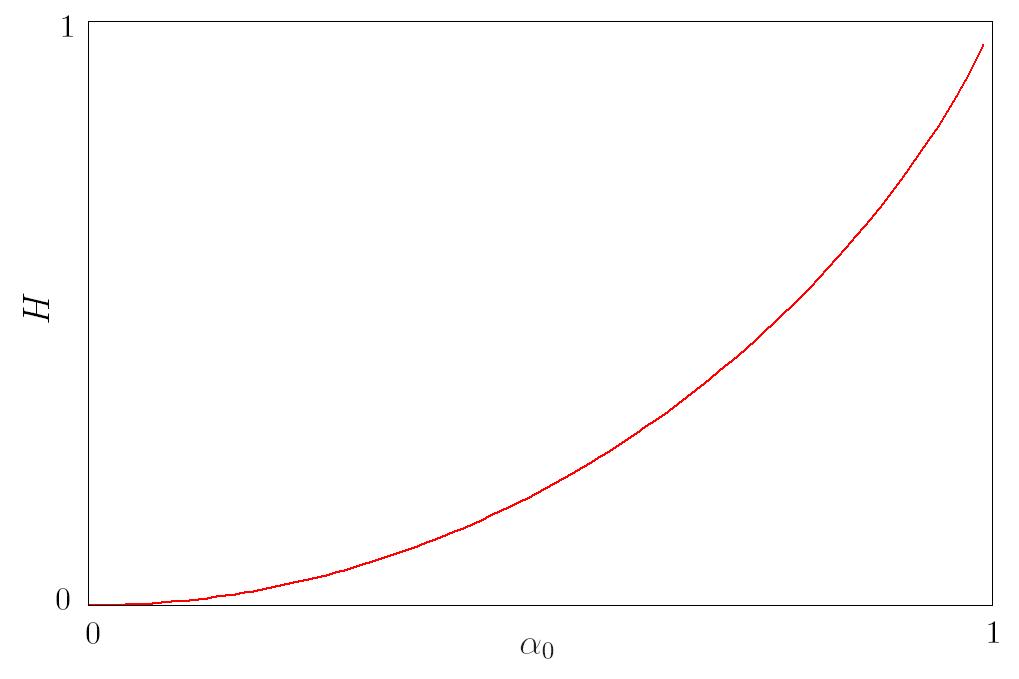}  
\caption{The energy $H(\alpha_{0})$ of the submaximal giant graviton in units of $\tfrac{16\pi^{2}T_{3}R^{4}}{9}$ (which is twice the energy of a dibaryon \cite{BHK}) as a function of its size $\alpha_{0}$.} \label{H-giant}
\end{center}
\end{figure}

This giant graviton is a BPS configuration, energetically degenerate with the point graviton, which exists by virtue of its motion along the $\chi_{1}$ angular direction.  It is dual to a subdeterminant operator of the form $\mathcal{O}_{n}(A_{1}B_{1})$.  The $\mathcal{R}$-charge $n$ maps to twice the angular momentum $2P_{\psi} = \tfrac{2}{3}(P_{\chi_{1}} + P_{\chi_{2}} + P_{\chi_{3}})$ along the fibre $\psi$, while the conformal dimension $\Delta$ maps to the energy $H$.  We now observe that the condition $H=P_{\chi_{1}}$, with $P_{\chi_{2}} = P_{\chi_{3}} = 0$, on the giant graviton is in agreement with $\Delta = \tfrac{3}{2}n$ for the dual operator\footnote{\label{geometry foot}It should be noted that in the gauge theory the only relevant charges are conformal dimension and $\mathcal{R}$-charge - the duals of energy and angular momentum. In particular, $\alpha_{0}$ makes no appearance. It is an open and extremely interesting question how the explicit geometry arises in \ref{radial dependence}.}.


\section{Fluctuation Analysis} \label{section - fluct}

For the purposes of the fluctuation analysis, it is convenient to describe the anti-de Sitter spacetime in terms of the coordinates $(t,v_{k})$ of \cite{DJM}:
\begin{equation}
ds_{AdS_{5}}^{2} = -\left(1 + \sum_{k}v_{k}^{2}\right)dt^{2} + \sum_{i,j}\left(\delta_{ij} - \frac{v_{i}v_{j}}{1 + \sum_{k}v_{k}^{2}}\right)dv_{i}dv_{j}.
\end{equation}

We shall also define $z_{i} \equiv \cos^{2}{\tfrac{\theta_{i}}{2}}$ in terms of which our orthogonal radial coordinates can be written as
\begin{equation}
\alpha = \sqrt{z_{1} + z_{2} - z_{1}z_{2}} ~~~~~~ \textrm{and} ~~~~~~ v = \frac{2\sqrt{z_{1}z_{2}}}{z_{1} + z_{2}} \equiv \sin{\beta},
~~~~~~ \textrm{with} ~~ \beta ~ \epsilon ~ \left[0,\tfrac{\pi}{2}\right].
\end{equation}
The surface of the giant graviton is described by $\alpha = \alpha_{0}$ constant, which is a shifted hyperbola in the $(z_{1},z_{2})$-plane:
\begin{equation}
\left(1 - z_{1}\right)\left(1 - z_{2}\right) = 1 - \alpha_{0}^{2}.
\end{equation}
The giant graviton solution can therefore be represented more simply in the coordinates $z_{i}$ (see figure \ref{fig - z1z2-plane} below), which shall also prove useful in our fluctuation analysis.
\begin{figure}[htb!]
\begin{center}
\includegraphics[width = 8.4cm, height = 7.5cm]{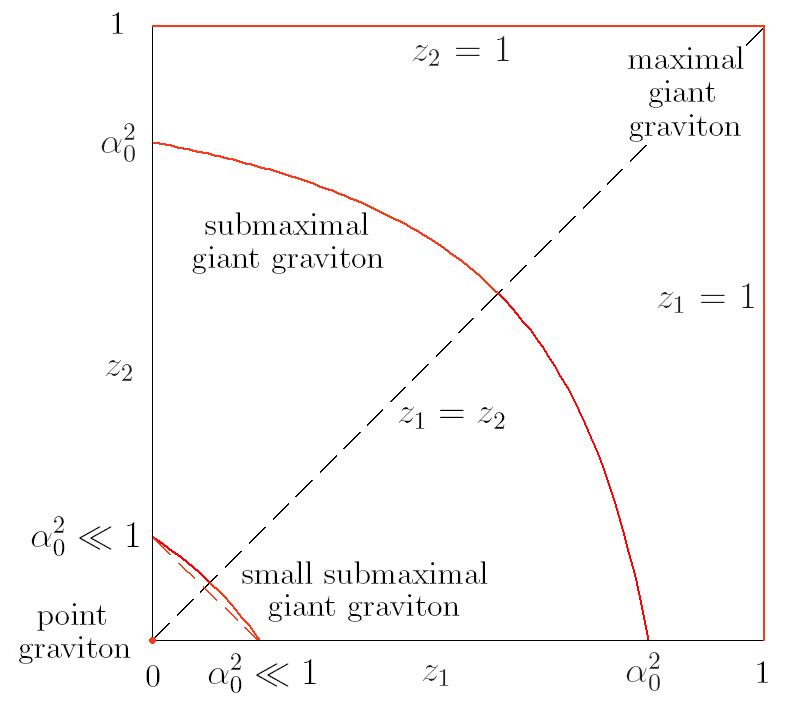}  
\caption{The giant graviton in the $(z_{1},z_{2})$-plane: the point graviton, small submaximal giant graviton (with approximate solution), submaximal giant graviton and maximal giant graviton (two dibaryons) are indicated in the sketch. The line $z_{1}=z_{2}$ (so $v=1$) separates the two regions which double-cover the $v ~ \epsilon ~ [0,1]$ interval.} \label{fig - z1z2-plane}
\end{center}
\end{figure}

We can invert these relations for $z_{1}$ and $z_{2}$ as follows:
\begin{eqnarray}
&& z_{1} \equiv \cos^{2}{\tfrac{\theta_{1}}{2}} = \frac{1}{v^{2}}\left(1-\sqrt{1-\alpha^{2}v^{2}}\right)\left(1 \mp \sqrt{1-v^{2}}\right) \\
&& z_{2} \equiv \cos^{2}{\tfrac{\theta_{2}}{2}} = \frac{1}{v^{2}}\left(1-\sqrt{1-\alpha^{2}v^{2}}\right)\left(1 \pm \sqrt{1-v^{2}}\right),
\end{eqnarray}
where the $\mp$ and $\pm$ distinguish between the regions $z_{1} \leq z_{2}$ and $z_{1} \geq z_{2}$.  Note that, on the first covering the $z_{i}$ may also be expressed in terms of the alternative radial worldvolume coordinate $\beta$:
\begin{eqnarray}
&& z_{1} = \frac{1}{\sin^{2}{\beta}}\left(1-\sqrt{1-\alpha^{2}\sin^{2}{\beta}}\right)\sin^{2}{\tfrac{\beta}{2}} \\
&& z_{2} = \frac{1}{\sin^{2}{\beta}}\left(1-\sqrt{1-\alpha^{2}\sin^{2}{\beta}}\right)\cos^{2}{\tfrac{\beta}{2}},
\end{eqnarray}
whereas we must interchange $\sin^{2}{\tfrac{\beta}{2}}$ and $\cos^{2}{\tfrac{\beta}{2}}$ on the second covering.  However, this is equivalent to taking $\beta \rightarrow \pi - \beta$, so both coverings may be parameterized by simply extending the range of $\beta$ to $[0,\pi]$.


\subsection{Fluctuation spectrum for the small submaximal giant graviton} \label{section - fluct - small giant}

A general analysis of small fluctuations about the submaximal giant graviton is presented in appendix \ref{appendix - general fluctuations}.  In order to solve the equations of motion resulting from the second order D3-brane action and obtain the fluctuation spectrum, we shall consider a small submaximal giant graviton with $\alpha_{0} \ll 1$.

It now becomes possible to approximate
\begin{eqnarray}
z_{1} \approx \alpha^{2}\sin^{2}{\tfrac{\beta}{2}} = \alpha^{2}(1-z) ~~~~~~ \textrm{and} ~~~~~~
z_{2} \approx \alpha^{2}\cos^{2}{\tfrac{\beta}{2}} = \alpha^{2}z,
\end{eqnarray}
where $z \equiv \cos^{2}{\tfrac{\beta}{2}}$ runs over the unit interval.  The second order D3-brane action \ref{action - second order general fluct} then simplifies as follows:
\begin{eqnarray}
& S \approx & \frac{T_{3}R^{4}\alpha_{0}^{2}\varepsilon^{2}}{16}\int dt ~ dz ~ d\chi_{2} ~ d\chi_{3} ~
\left\{\sum_{k}\left[-\frac{3}{2}\delta v_{k}^{2} - \left(\tilde{\p}~\delta v_{k}\right)^{2}\right] \right. \\
\nonumber && \left. \hspace{2.0cm} - \frac{2}{3}\left(\tilde{\p}~\delta\alpha\right)^{2}
- \frac{2}{3}\left(\tilde{\p}~\tilde{\delta\chi}_{1}\right)^{2}
+ \frac{4}{3}\left[3\dot{\tilde{\delta\chi}}_{1} - \left(\p_{\chi_{2}}\tilde{\delta\chi}_{1}\right) - \left(\p_{\chi_{3}}\tilde{\delta\chi}_{1}\right)\right]\delta\alpha\right\},
\end{eqnarray}
with $\tilde{\delta\chi}_{1} \equiv \tfrac{1}{\alpha_{0}}~\delta\chi_{1}$ and $\tilde{\p} \equiv \alpha_{0}\p$ rescaled for convenience. The (spacetime) gradient squared is then given by
\begin{eqnarray}
\nonumber & \left(\tilde{\p}f\right)^{2} \equiv & 6z(1-z)\left(\p_{z}f\right)^{2} - \tfrac{3}{2}\dot{f}^{2} + \tfrac{5}{2}\left(\p_{\chi_{2}}f\right)^{2} + \tfrac{5}{2}\left(\p_{\chi_{3}}f\right)^{2} + \dot{f}\left(\p_{\chi_{2}}f\right) + \dot{f}\left(\p_{\chi_{3}}f\right) \\
&& - \tfrac{1}{3}\left(\p_{\chi_{2}}f\right)\left(\p_{\chi_{3}}f\right) + \tfrac{8}{3}\tfrac{z}{\left(1-z\right)}\left(\p_{\chi_{2}}f\right)^{2}
+ \tfrac{8}{3}\tfrac{\left(1-z\right)}{z}\left(\p_{\chi_{3}}f\right)^{2}.
\end{eqnarray}

The equations of motion are
\begin{eqnarray}
&& \tilde{\Box}\delta v_{k} - \tfrac{3}{2}\delta v_{k} = 0 \\
&& \tilde{\Box}\delta y_{\pm} \mp 3i\dot{\delta y_{\pm}} \pm i\left(\p_{\chi_{2}}\delta y_{\pm}\right)\pm i\left(\p_{\chi_{3}}\delta y_{\pm}\right) = 0,
\end{eqnarray}
where we define $\delta y_{\pm} \equiv \delta\alpha \pm i\tilde{\delta\chi}_{1}$.  The (rescaled) d'Alembertian on the worldvolume of the giant graviton is
\begin{eqnarray} \label{dAlem}
& \tilde{\Box} \equiv & 6\p_{z}\left\{z\left(1-z\right)\p_{z}\right\} \\
\nonumber && - \tfrac{3}{2}~\p_{t}^{2} + \tfrac{5}{2}~\p_{\chi_{2}}^{2} + \tfrac{5}{2}~\p_{\chi_{3}}^{2} + \p_{t}\p_{\chi_{2}} + \p_{t}\p_{\chi_{3}} - \tfrac{1}{3}~\p_{\chi_{2}}\p_{\chi_{3}}
 + \tfrac{8}{3}\tfrac{z}{\left(1-z\right)}~\p_{\chi_{2}}^{2} + \tfrac{8}{3}\tfrac{\left(1-z\right)}{z}~\p_{\chi_{3}}^{2}.
\end{eqnarray}

We shall now expand these fluctuations as
\begin{eqnarray}
&& \delta v_{k}(t,z,\chi_{2},\chi_{3}) = \sum_{m,n,s} C_{smn} \Psi_{smn}(t,z,\chi_{2},\chi_{3}) \\
&& \delta y_{\pm}(t,z,\chi_{2},\chi_{3}) = \sum_{m,n,s} \tilde{C}_{smn} \Psi_{smn}(t,z,\chi_{2},\chi_{3}),
\end{eqnarray}
in terms of the eigenfunctions $\Psi_{smn}(t,z,\chi_{2},\chi_{3})$ described in appendix \ref{appendix - evp - small giant}.  Insisting that the equations of motion must be satisfied places the following constraints on the frequencies $\omega_{smn}$ (already contained in the definition
\ref{eigenfunctions - small giant} of these eigenfunctions):
\begin{eqnarray}
&& \left[\omega^{k}_{smn} + \tfrac{1}{4}\left(m + n\right)\right]^{2} = 4l\left(l+1\right) + 1 \\
&& \left[\omega^{\pm}_{smn} + \tfrac{1}{4}\left(m + n\right) \mp 1\right]^{2} = 4l\left(l+1\right) + 1,
\end{eqnarray}
with $l \equiv s + \max{\left\{\tfrac{1}{2}|m+n|,\tfrac{1}{2}|m-n|\right\}}$.  Here $s$, $m$ and $n$ are integers, with $s$ non-negative.

Notice that the above expressions are always positive, indicating that the frequencies $\omega^{k}_{smn}$ and $\omega^{\pm}_{smn}$ are real.  Hence the small submaximal giant graviton is a stable configuration.  Furthermore, in this approximation $\alpha_{0} \ll 1$, the fluctuation spectrum is independent of the size of the giant graviton $\alpha_{0}$.  This appears not to be true in general, however, as we shall now observe by comparing these results with the fluctuation spectrum of the maximal giant graviton.


\subsection{Fluctuation spectrum for the maximal giant graviton} \label{section - fluct - max giant}

\textbf{\emph{Dibaryon spectrum}}

Let us briefly review the construction \cite{BHK} of the spectrum of small fluctuations about each of the dibaryons (described in figure \ref{dual dibaryons} below).  These correspond to the determinant operators $\det{A_{1}}$ and $\det{B_{1}}$ respectively in the dual gauge theory.
\begin{figure}[htb!]
\begin{center}
\centering
\includegraphics[width = 12.5cm, height = 6.5cm]{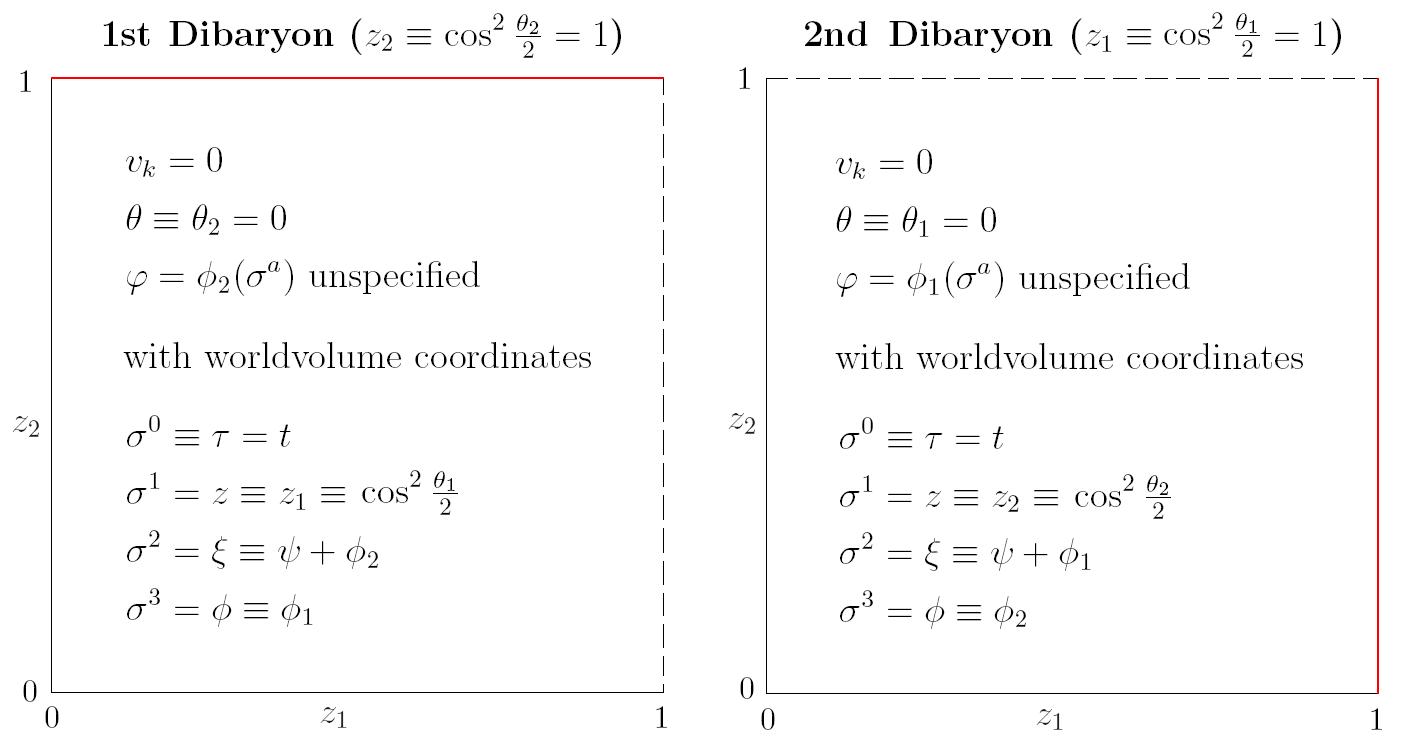} 
\caption{A description of the two dibaryons defined by $z_{2}=1$ and $z_{1}=1$ respectively, which together make up the maximal giant graviton. These are D3-branes wrapped on one of the 2-spheres and the (shifted) fibre direction.} \label{dual dibaryons}
\end{center}
\end{figure}

The D3-brane action, which describes one of these dibaryons is given by
\begin{equation}
S_{(0)} = -\frac{T_{3}R^{4}}{9}\int_{\Sigma} d^{4}\sigma \left\{1 - \frac{4}{3}\dot{\varphi}(\sigma^{a})\right\}
\end{equation}
with $T_{3}=\tfrac{1}{(2\pi)^{3}}$ the tension.  The momentum conjugate to $\varphi$ and the energy are
\begin{equation}
P_{\varphi} = \frac{32\pi^{2}T_{3}R^{4}}{27} = \frac{4R^{4}}{27\pi} = N
~~~~~~~~ \textrm{and} ~~~~~~~~
H = 8\pi^{2}T_{3}R^{4} = \frac{R^{4}}{9\pi} = \frac{3}{4}N.
\end{equation}
Here the energy is simply the volume of the wrapped D3-brane which may be identified with the conformal dimension $\Delta$ of the corresponding dibaryon operator in Klebanov-Witten theory.

Let us now consider small fluctuations about a dibaryon. Our ansatz is as follows:
\begin{equation}
v_{k} = \varepsilon\delta v_{k}(\sigma^{a}) ~~~~~~ \textrm{and} ~~~~~~ \theta = \varepsilon\delta\theta(\sigma^{a}), ~~~~~~ \textrm{with} ~~~
\varphi(\sigma^{a}) ~~ \textrm{unspecified}.
\end{equation}
Here $\sigma^{a} = (t,z,\xi,\phi)$ are the worldvolume coordinates and $\varepsilon$ is a small parameter.  We shall also define $\delta y_{1} \equiv \delta\theta \cos{\varphi}$ and $\delta y_{2} \equiv \delta\theta \sin{\varphi}$.

The D3-brane action is then $S \approx S_{(0)} + \varepsilon^{2}S_{(2)}$, with $S_{(0)}$ the original action for the dibaryon and
\begin{eqnarray}
\nonumber && \!\! S_{(2)} = -\frac{T_{3}R^{4}}{9}\int_{\Sigma} d^{4}\sigma
\left\{ \sum_{k}\left[\delta v_{k}^{2} - \dot{\delta v}_{k}^{2} + \left(\nabla\delta v_{k}\right)^{2}\right]
+ \frac{1}{6}\sum_{i}\left[-\dot{\delta y_{i}}^{2} + \left(\nabla\delta y_{i}\right)^{2}\right] \right. \\
&& \hspace{4.0cm} \left. - \frac{2}{3}\left[\delta y_{2}\dot{\delta y_{1}} - \delta y_{1}\dot{\delta y_{2}}\right]
+ \left[\delta y_{2}\left(\p_{\xi} \delta y_{1}\right) - \delta y_{1}\left(\p_{\xi} \delta y_{2}\right)\right] \right\} \hspace{1.2cm}
\end{eqnarray}
the second order corrections.  The gradient squared of any function $f(z,\xi,\phi)$ on the worldvolume of the dibaryon is
\begin{eqnarray}
\nonumber & \left(\nabla f\right)^{2} \equiv & 6z(1-z)(\p_{z}f)^{2} \\
&& + \tfrac{3}{2}\tfrac{1}{z(1-z)}\left\{\left[2z(1-z)+1\right](\p_{\xi}f)^{2} + (\p_{\phi}f)^{2} - 2(2z-1)(\p_{\xi}f)(\p_{\phi}f)\right\}. \hspace{1.0cm}
\end{eqnarray}
The equations of motion then take the form
\begin{eqnarray}
&& \delta v_{k} + \ddot{\delta v}_{k} - \nabla^{2}\delta v_{k} = 0 \\
&& \ddot{\delta y}_{\pm} - \nabla^{2} \delta y_{\pm} \mp 4i\dot{\delta y}_{\pm} \pm 6i\left(\p_{\xi} \delta y_{\pm}\right) = 0,
\end{eqnarray}
with $\delta y_{\pm} \equiv \delta y_{1} \pm i \delta y_{2} = \delta\theta~e^{\pm i\varphi}$ in terms of which the latter equations decouple.  Here the Laplacian is given by
\begin{equation} \label{Laplacian}
\nabla^{2} \equiv 6\p_{z}\left\{z(1-z)\p_{z}\right\} +
\tfrac{3}{2}\tfrac{1}{z(1-z)}\left\{\left[2z(1-z)+1\right]\p_{\xi}^{2} + \p_{\phi}^{2} - 2(2z-1)\p_{\xi}\p_{\phi}\right\}.
\end{equation}

Expanding the fluctuations in terms of their Fourier modes
\begin{eqnarray}
&& \delta v_{k}(t,z,\xi,\phi) = \sum_{m,n,s} C_{smn} e^{-i\omega^{k}_{smn}t} \Phi_{smn}(z,\xi,\phi) \\
&& \delta y_{\pm}(t,z,\xi,\phi) = \sum_{m,n,s} \tilde{C}_{smn} e^{-i\omega^{\pm}_{smn}t} \Phi_{smn}(z,\xi,\phi),
\end{eqnarray}
where $\Phi_{smn}(z,\xi,\phi)$ are the eigenfunctions \ref{eigenfunctions - dibaryon} of the stationary eigenvalue problem, we obtain the following spectrum for one of the dibaryons \cite{BHK}:
\begin{eqnarray}
&& \left(\omega^{k}_{smn}\right)^{2} = 6l(l+1) + 3m^{2} + 1 \label{omegak} \\
&& \left(\omega^{\pm}_{smn} \pm 2\right)^{2} =  6l(l+1) + 3(m \mp 1)^{2} + 1, \label{omegapm}
\end{eqnarray}
with $l = s + \max\left\{|m|,|n|\right\}$.  Here $s$ and $n$ are integers, with $s$ non-negative, and $m$ is an integer or half-integer.

\bigskip

\textbf{\emph{Spectrum of the maximal giant graviton}}

Small fluctuations about the maximal giant graviton cannot be described by our general ansatz \ref{fluctuation ansatz} with $\alpha_{0} = 1$. We must now require
\begin{equation}
\sin^{2}{\tfrac{\theta_{1}}{2}}\sin^{2}{\tfrac{\theta_{2}}{2}} = (1-z_{1})(1-z_{2}) = 1-\alpha^{2} = \rho^{2}
\end{equation}
to be of $O\left(\varepsilon^{2}\right)$. Hence we shall modify our ansatz as follows:
\begin{equation}
v_{k} = \varepsilon\delta v_{k}(\sigma^{a}) ~~~~~~~~ \textrm{and} ~~~~~~~~ \rho = \varepsilon\delta\rho(\sigma^{a}),
\end{equation}
with worldvolume coordinates $\sigma = (t,z,\xi,\phi)$ covering both halves of the maximal giant graviton (dibaryons) on which $\phi_{1}(\sigma^{a})$ and $\phi_{2}(\sigma^{a})$ respectively remain unspecified.

The fluctuations of the anti-de Sitter coordinates are simply the sum of these fluctuation about each dibaryon:
\begin{equation}
\delta v_{k} = \sum_{s,m,n} e^{-i\omega^{k}_{smn}t} \left\{ C^{(1)}_{smn} \Phi_{smn}(z_{1},\psi+\phi_{2},\phi_{1}) + C^{(2)}_{smn}, \Phi_{smn}(z_{2},\psi+\phi_{1},\phi_{2})  \right\},
\end{equation}
where $\Phi_{smn}$ are the eigenfunctions \ref{eigenfunctions - dibaryon} of the Laplacian on a dibaryon and the frequencies satisfy \ref{omegak}. We must impose the condition $C^{(1)}_{smm} = C^{(2)}_{smm}$ for the fluctuations which do not vanish at $z_{1} = z_{2} = 1$ to match up.

Now, the fluctuations of the radial $T^{1,1}$ coordinate $\rho$ can be written as
\begin{equation}
\delta \rho = \tfrac{1}{2}\left(1-z_{2}\right)^{\frac{1}{2}}\delta\theta_{1}(t,z_{2},\psi+\phi_{1},\phi_{2}) + \tfrac{1}{2}\left(1-z_{1}\right)^{\frac{1}{2}}\delta\theta_{2}(t,z_{1},\psi+\phi_{2},\phi_{1}).
\end{equation}
However, we must allow, not only for the usual $\delta\theta$ fluctuations of the dibaryon, but also for the possibility that $\delta\theta$ diverges like $(1-z)^{-\frac{1}{2}}$ as $z$ goes to $1$ (see appendix \ref{appendix - evp - dibaryon} for details).  These yield non-vanishing, but finite, contributions to $\delta \rho$ at $z_{1}=z_{2}=1$ (which must match). These additional fluctuations correspond to open strings stretched between the two halves of the maximal giant graviton.

The fluctuations $\delta y^{(1)}_{\pm} \equiv \delta\theta_{2}~e^{\pm i \phi_{2}}$ and $\delta y^{(2)}_{\pm} \equiv \delta\theta_{1}~e^{\pm i \phi_{1}}$ of the $T^{1,1}$ coordinates transverse to the different dibaryons, which both contribute to $\delta\rho$, are then
\begin{eqnarray}
\nonumber & \delta y^{(1)}_{\pm} =
& \sum_{s,m,n} \tilde{C}^{(1)}_{smn} ~ e^{-i\omega^{\pm}_{smn}t} ~ \Phi_{smn}(z_{1},\psi+\phi_{2},\phi_{1}) \\
&& + \sum_{^{s,m,}_{n=m\pm 1}} \tilde{C}^{\textrm{mod}(1)}_{smn} ~ e^{-i\omega^{\textrm{mod},\pm}_{smn}t} ~ \Phi^{\textrm{mod}}_{smn}(z_{1},\psi+\phi_{2},\phi_{1})
\end{eqnarray}
and
\begin{eqnarray}
\nonumber & \delta y^{(2)}_{\pm} =
& \sum_{s,m,n} \tilde{C}^{(2)}_{smn} ~ e^{-i\omega^{\pm}_{smn}t} ~ \Phi_{smn}(z_{2},\psi+\phi_{1},\phi_{2}) \\
&& + \sum_{^{s,m,}_{n=m\pm 1}} \tilde{C}^{\textrm{mod}(2)}_{smn} ~ e^{-i\omega^{\textrm{mod},\pm}_{smn}t} ~ \Phi^{\textrm{mod}}_{smn}(z_{2},\psi+\phi_{1},\phi_{2}),
\end{eqnarray}
where $\Phi^{\textrm{mod}}_{smn}$ are the modified eigenfunctions \ref{eigenfunction - modified dibaryon} and we impose the matching condition $\tilde{C}^{\textrm{mod}(1)}_{smn} = \tilde{C}^{\textrm{mod}(2)}_{smn}$. The frequencies $\omega^{\pm}_{smn}$ of the original contributions still satisfy \ref{omegapm}, while the modified frequencies $\omega^{\textrm{mod},\pm}_{smn}$ satisfy an identical condition, but with $l^{\textrm{mod}} = s + \tfrac{1}{2}(|m+n| - 1)$.  Since these are non-negative integers (the unmodified $l$ values could be integer or half-integer), the spectrum of the maximal giant graviton is entirely contained within the original spectrum of the separate dibaryons.  The frequencies are therefore still real, indicating stability.


\section{Open strings excitations} \label{section - excitations}

We shall now turn our attention to open string excitations of the submaximal giant graviton.  Since a full quantum description of strings in $AdS_{5}\times T^{1,1}$ remains unknown, we shall consider a simplifying limit \cite{Berenstein-et-al}: short open strings moving in the pp-wave geometry\footnote{Penrose limits of the $AdS_{5}\times T^{1,1}$ background were studied in \cite{T11 Penrose limits}.} associated with a null geodesic on the worldvolume of the giant graviton. (Note that different null geodesics produce distinct results, due to the non-spherical nature of the submaximal giant, and we shall discuss two possibilities.)


\subsection{Short open pp-wave strings: endpoints on the null geodesic parameterized by $t = \chi_{1} = \chi_{2} \equiv u$}
\label{section - excitations - null2}

Let us consider the null geodesic
\begin{equation}
t = \chi_{1} = \chi_{2} = u, ~~~~~~ v_{k} = 0, ~~~~~~ \alpha = \alpha_{0} ~~~~~~ \textrm{and} ~~~~~~ v = 0,
\end{equation}
on the worldvolume of the submaximal giant graviton with size $\alpha_{0}$.  We observe that $\theta_{1} = 2\arccos{\alpha_{0}}$ and $\theta_{2} = \pi$ then specifies the trajectory on the two 2-spheres.

To construct the pp-wave background, we choose the ansatz
\begin{eqnarray}
\nonumber && t = u + \frac{\xi}{R^{2}} \hspace{5.65cm} v_{k} = \frac{x_{k}}{R} \\
\nonumber && \chi_{1} = u - \frac{\xi}{R^{2}} - \sqrt{\frac{2}{3}}\frac{\alpha_{0}}{\sqrt{1-\alpha_{0}^{2}}}~\frac{y_{1}}{R} \hspace{2.0cm}
\alpha = \alpha_{0} + \sqrt{\frac{3}{2}}\sqrt{1-\alpha_{0}^{2}}~\frac{y_{2}}{R} \\
\nonumber && \chi_{2} = u - \frac{\xi}{R^{2}} + \sqrt{\frac{2}{3}}\frac{\left(2-\alpha_{0}^{2}\right)}{\alpha_{0}\sqrt{1-\alpha_{0}^{2}}} ~\frac{y_{1}}{R} \hspace{1.6cm}
v = \frac{\sqrt{6}}{\alpha_{0}} \frac{\tilde{y}}{R}\\
&& \chi_{3} = \frac{4}{3}\chi,
\end{eqnarray}
which corresponds to setting
\begin{equation}
\theta_{1} = 2\arccos\alpha_{0} - \sqrt{6}~\frac{y_{2}}{R} ~~~~~~ \textrm{and} ~~~~~~ \theta_{2} = \pi - \sqrt{6}~\frac{\tilde{y}}{R},
\end{equation}
with $\tilde{y}_{1} = \tilde{y}\cos{\chi}$ and $\tilde{y}_{2} = \tilde{y}\sin{\chi}$.  Now, taking the Penrose limit, in which $R$ becomes large and we zoom in on this null geodesic, we obtain the pp-wave metric
\begin{eqnarray}
\nonumber & ds^{2} = & -4dud\xi - \left\{\sum_{k=1}^{4}{x_{k}^{2}} + \frac{15}{16}\sum_{i=1}^{2}{\tilde{y}_{i}^{2}}\right\}du^{2} + \sum_{k=1}^{4}{dx_{k}^{2}} + \sum_{i=1}^{2}{dy_{i}^{2}} + \sum_{i=1}^{2}{d\tilde{y}_{i}^{2}} \\
&& + 4y_{2}dy_{1}du + \tfrac{1}{2}\left(\tilde{y}_{1}d\tilde{y}_{2}-\tilde{y}_{2}d\tilde{y}_{1}\right)du,
\end{eqnarray}
which has a flat direction $y_{1}$ (for a discussion of isometries in pp-wave limits, see \cite{Grignani:2009ny}). The 5-form field strength, in this pp-wave limit, becomes constant:
\begin{equation}
F_{5} = 4 du \wedge dx_{1} \wedge dx_{2} \wedge dx_{3} \wedge dx_{4} + 8 du \wedge dy_{1} \wedge dy_{2} \wedge d\tilde{y}_{1} \wedge d\tilde{y}_{2}.
\end{equation}

In the lightcone gauge $u=2p^{u}\tau$, the bosonic part of the Polyakov action for open strings moving in this pp-wave geometry is
\begin{eqnarray}
\nonumber && S = \int{d\tau}\int{\frac{d\sigma}{2\pi}} \left\{\sum_{K=1}^{4}\left(\tfrac{1}{2}\dot{X}_{K}^{2} - \tfrac{1}{2}\left(X'_{K}\right)^{2} - \tfrac{1}{2}m^{2}X_{K}^{2}\right) + \sum_{I=1}^{2}\left(\tfrac{1}{2}\dot{Y}_{I}^{2} - \tfrac{1}{2}\left(Y'_{I}\right)^{2}\right) + 2mY_{2}\dot{Y}_{1} \right. \\
&& \left. \hspace{3.2cm} + \sum_{I=1}^{2}\left(\tfrac{1}{2}\dot{\tilde{Y}}_{I}^{2} - \tfrac{1}{2}\left(\tilde{Y}'_{I}\right)^{2} - \tfrac{15}{32}m^{2}\tilde{Y}_{I}^{2}\right) + \tfrac{1}{4}m\left(\tilde{Y}_{1}\dot{\tilde{Y}}_{2}-\tilde{Y}_{2}\dot{\tilde{Y}}_{1}\right)\right\},
\end{eqnarray}
with $m \equiv 2p^{u}$. Notice that the equations of motion for each pair of embedding coordinates $Y_{I}$ and $\tilde{Y}_{I}$ decouple, if we define
\begin{equation}
Y_{\pm} \equiv \frac{1}{\sqrt{2}}\left(Y_{1} \pm iY_{2}\right) ~~~~~~ \textrm{and} ~~~~~~
\tilde{Y}_{\pm} \equiv \frac{1}{\sqrt{2}}\left(\tilde{Y}_{1} \pm i\tilde{Y}_{2}\right).
\end{equation}
The assumption that the open pp-wave string ends on the submaximal giant graviton (which becomes a flat D3-brane in the large $R$ limit) implies that the $X_{K}$ and $Y_{I}$ satisfy Dirichlet boundary conditions, whereas the $\tilde{Y}_{I}$ must obey Neumann boundary conditions. Quantizing the open string embedding coordinates, we then obtain
\begin{eqnarray}
&& \!\!\! X_{K}\left(\tau,\sigma\right) = \sum_{n=1}^{\infty}\sqrt{\frac{2}{\omega_{n}}}
\left\{\alpha_{n}^{K}e^{-i\omega_{n}\tau} + \left(\alpha_{n}^{K}\right)^{\dag}e^{i\omega_{n}\tau}\right\}\sin{\left(n\sigma\right)} \\
&& \!\!\! Y_{\pm}\left(\tau,\sigma\right) = \sum_{n=1}^{\infty}\sqrt{\frac{2}{\omega_{n}}}
\left\{\beta^{\mp}_{n}e^{-i\omega^{\mp}_{n}\tau} + \left(\beta^{\pm}_{n}\right)^{\dag}e^{i\omega^{\pm}_{n}\tau}\right\}\sin{\left(n\sigma\right)} \\
\nonumber && \!\!\! \tilde{Y}_{\pm}\left(\tau,\sigma\right) =
\frac{1}{\sqrt{m}}\left\{\tilde{\beta}^{\pm}_{n}e^{-i\tilde{\omega}^{\pm}_{0}\tau} + \left(\tilde{\beta}^{\mp}_{n}\right)^{\dag}e^{i\tilde{\omega}^{\mp}_{0}\tau}\right\}
+ \sum_{n=1}^{\infty}\sqrt{\frac{2}{\omega_{n}}}\left\{\tilde{\beta}^{\pm}_{n}e^{-i\tilde{\omega}^{\pm}_{n}\tau} + \left(\tilde{\beta}^{\mp}_{n}\right)^{\dag}e^{i\tilde{\omega}^{\mp}_{n}\tau}\right\} \cos{\left(n\sigma\right)}, \\
\end{eqnarray}
where we define $\omega^{\pm}_{n} \equiv \omega_{n} \pm m$ and $\tilde{\omega}^{\pm}_{n} \equiv \omega_{n} \pm \tfrac{1}{4}m$ in terms of $\omega_{n} \equiv \sqrt{m^{2}+n^{2}}$.  The creation and annihilation operators satisfy the following commutation relations:
\begin{equation} \label{comm-relns}
\left[\alpha_{n}^{K_{1}},\left(\alpha_{l}^{K_{2}}\right)^{\dag}\right] = \delta^{K_{1}K_{2}}\delta_{nl} ~~~~ \textrm{and} ~~~~
\left[\beta^{\pm}_{n},\left(\beta^{\pm}_{l}\right)^{\dag}\right]
= \left[\tilde{\beta}^{\pm}_{n},\left(\tilde{\beta}^{\pm}_{l}\right)^{\dag}\right] = \delta_{nl},
\end{equation}
with all others zero. The lightcone Hamiltonian $H_{lc} = \tfrac{1}{m}H$ is quadratic in the embedding coordinates, and can be written in terms of these (normal ordered) harmonic oscillators:
\begin{eqnarray}
\nonumber && H_{lc} = \sum_{n=1}^{\infty}\sum_{K=1}^{4}
\frac{\omega_{n}}{m}\left(\alpha_{n}^{K}\right)^{\dag}\alpha_{n}^{K}
+ \sum_{n=1}^{\infty}\left\{\frac{\omega^{+}_{n}}{m}\left(\beta_{n}^{+}\right)^{\dag}\beta_{n}^{+} + \frac{\omega^{-}_{n}}{m}\left(\beta_{n}^{-}\right)^{\dag}\beta_{n}^{-}\right\} \\
&& ~ \hspace{0.85cm} + \sum_{n=0}^{\infty}\left\{\frac{\tilde{\omega}^{+}_{n}}{m}\left(\tilde{\beta}_{n}^{+}\right)^{\dag}\tilde{\beta}_{n}^{+} + \frac{\tilde{\omega}^{-}_{n}}{m}\left(\tilde{\beta}_{n}^{-}\right)^{\dag}\tilde{\beta}_{n}^{-}\right\}
\end{eqnarray}
which leads to a spectrum for $H_{lc}$\footnote{Note that in this Penrose limit there are no zero energy solutions, but in the limit where $m \to \infty$, the $\beta^-$ modes look like massless excitations with no zero mode.}
\begin{eqnarray}
\label{pp-wave hamiltonian 1-alpha}
& \alpha_{n}^{K}: & ~~ \frac{\omega_{n}}{m} = \sqrt{1 + \frac{n^{2}}{m^{2}}}
\hspace{3cm} \textrm{with } n = 1, 2, \ldots \hspace{1.8cm} \\
\label{pp-wave hamiltonian 1-beta}
& \beta_{n}^{\pm}: & ~~ \frac{\omega^{\pm}_{n}}{m} = \sqrt{1 + \frac{n^{2}}{m^{2}}} \pm 1
\hspace{2.2cm} \textrm{with } n = 1, 2, \ldots \hspace{1.8cm} \\
\label{pp-wave hamiltonian 1-beta-tilde}
& \tilde{\beta}_{n}^{\pm}: & ~~ \frac{\tilde{\omega}^{\pm}_{n}}{m} = \sqrt{1 + \frac{n^{2}}{m^{2}}} \pm \frac{1}{4}
\hspace{2.1cm} \textrm{with } n = 0, 1, 2, \ldots \hspace{1.8cm}
\end{eqnarray}

Let us now consider the interpretation of these results in terms of the original AdS coordinates.  In the large $R$ limit, we assume that physical momenta are finite.  This means that the lightcone momentum
\begin{equation} \label{mass null2}
p_{\xi} = -\frac{1}{R^{2}}\left(E + P_{\chi_{1}} + P_{\chi_{2}}\right) = -m \;,
\end{equation}
the lightcone Hamiltonian
\begin{equation}
H_{lc} = -p_{u} = E - P_{\chi_{1}} - P_{\chi_{2}},
\end{equation}
and the momenta
\begin{equation}
p_{y_{1}} = \sqrt{\frac{2}{3}}\frac{1}{R}
\frac{\left[-\alpha_{0}^{2}P_{\chi_{1}} + \left(2-\alpha_{0}^{2}\right)P_{\chi_{2}}\right]}{\alpha_{0}\sqrt{1-\alpha_{0}^{2}}}
~~~~~~ \textrm{and} ~~~~~~ p_{\chi} = \frac{4}{3}P_{\chi_{3}},
\end{equation}
are all finite.  Here $E = i\partial_t$ and $P_{i} = -i\partial_{i}$ ($p_{i} = -i\partial_{i}$) are the $AdS_{5}\times T^{1,1}$ (pp-wave) momenta.  With these assumptions, we observe that
\begin{equation} \label{charges pp-wave null2}
E = J_{\chi_1} + P_{\chi_2} + O(1), ~~~~~~ P_{\chi_{2}} = \frac{\alpha_0^2}{2-\alpha_0^2} \hspace{0.1cm} P_{\chi_1} + O(R),  ~~~~~~ P_{\chi_{3}} = O(1)
\end{equation}
and $P_{\chi_1}$ should be of $O(R^{2})$ for the `mass' \ref{mass null2} to be non-vanishing.


\subsection{Short open pp-wave strings: endpoints on the null geodesic parameterized by $t = \chi_{1} = \chi_{+} \equiv u$}
\label{section - excitations - null+}

We shall now consider the null geodesic\footnote{Note that the choice $v=1$ is not necessary to obtain a null geodesic, but is required
for the associated pp-wave background to be consistent.}
\begin{equation}
t = \chi_{1} = \chi_{+} = u, ~~~~~~ v_{k} = 0, ~~~~~~ \alpha = \alpha_{0}, ~~~~~~ v = 1 ~~~~~ \textrm{and} ~~~~~ \chi_{-} = 0,
\end{equation}
with $\chi_{\pm} \equiv \tfrac{1}{2}\left(\chi_{2} \pm \chi_{3} \right)$. Here also $\theta_{1} = \theta_{2} = 2\arcsin{\left(1-\alpha_{0}^{2}\right)^{1/4}} \equiv \theta_{0}$ and we notice that this setup is symmetric under interchange of the 2-sphere coordinates.

The ansatz for the pp-wave background is then given by
\begin{eqnarray}
\nonumber && \!\!\! t = u + \frac{\xi}{R^{2}} \hspace{5.95cm} v_{k} = \frac{x_{k}}{R}\\
\nonumber && \!\!\! \chi_{1} = u - \frac{\xi}{R^{2}} - \frac{2}{\sqrt{3}}\frac{\sqrt{1-\sqrt{1-\alpha_{0}^{2}}}}{\left(1-\alpha_{0}^{2}\right)^{1/4}}~\frac{y_{1}}{R} \hspace{1.2cm}
\alpha = \alpha_{0} + \frac{\sqrt{3}}{\alpha_{0}}\left(1-\alpha_{0}^{2}\right)^{3/4}\sqrt{1-\sqrt{1-\alpha_{0}^{2}}}~\frac{y_{2}}{R} \\
\nonumber && \!\!\! \chi_{+} = u - \frac{\xi}{R^{2}} + \frac{2}{\sqrt{3}}\frac{\left(1-\alpha_{0}^{2}\right)^{1/4}}{\sqrt{1-\sqrt{1-\alpha_{0}^{2}}}}~\frac{y_{1}}{R} \hspace{1.15cm}
v = 1 - \frac{3}{2}\frac{\sqrt{1-\alpha_{0}^{2}}}{\left(1 - \sqrt{1-\alpha_{0}^{2}}\right)}~\left(\frac{\tilde{y}_{2}}{R}\right)^{2} \\
&& \!\!\! \chi_{-} = \frac{2}{\sqrt{3}}\frac{1}{\left(1-\alpha_{0}^{2}\right)^{1/4}\sqrt{1-\sqrt{1-\alpha_{0}^{2}}}}~\frac{\tilde{y}_{1}}{R},
\end{eqnarray}
which corresponds to choosing
\begin{equation}
\theta_{1} = \theta_{0} -\sqrt{3}\left(\frac{y_{2}}{R} + \frac{\tilde{y}_{2}}{R}\right) ~~~~~~ \textrm{and} ~~~~~~
\theta_{2} = \theta_{0} -\sqrt{3}\left(\frac{y_{2}}{R} - \frac{\tilde{y}_{2}}{R}\right).
\end{equation}
Again, we take the large $R$ Penrose limit to obtain the pp-wave geometry
\begin{equation}
ds^{2} = -4dud\xi - \left(\sum_{k=1}^{4}{x_{k}^{2}}\right)du^{2} + \sum_{k=1}^{4}{dx_{k}^{2}} + \sum_{i=1}^{2}{dy_{i}^{2}} + \sum_{i=1}^{2}{d\tilde{y}_{i}^{2}} + 4y_{2}dy_{1}du + 4\tilde{y}_{2}d\tilde{y}_{1}du,
\end{equation}
with flat directions $y_{1}$ and $\tilde{y}_{1}$. The 5-form field strength becomes
\begin{equation}
F_{5} = 4 du \wedge dx_{1} \wedge dx_{2} \wedge dx_{3} \wedge dx_{4} - 4 du \wedge dy_{1} \wedge dy_{2} \wedge d\tilde{y}_{1} \wedge d\tilde{y}_{2}.
\end{equation}

The bosonic part of the Polyakov action for open strings in the lightcone gauge is
\begin{eqnarray}
\nonumber && S = \int{d\tau}\int{\frac{d\sigma}{2\pi}} \left\{\sum_{K=1}^{4}\left(\tfrac{1}{2}\dot{X}_{K}^{2} - \tfrac{1}{2}\left(X'_{K}\right)^{2} - \tfrac{1}{2}m^{2}X_{K}^{2}\right) + \sum_{I=1}^{2}\left(\tfrac{1}{2}\dot{Y}_{I}^{2} - \tfrac{1}{2}\left(Y'_{I}\right)^{2}\right) + 2mY_{2}\dot{Y}_{1} \right. \\
&& \left. \hspace{3.2cm} + \sum_{I=1}^{2}\left(\tfrac{1}{2}\dot{\tilde{Y}}_{I}^{2} - \tfrac{1}{2}\left(\tilde{Y}'_{I}\right)^{2}\right) + 2m\tilde{Y}_{2}\dot{\tilde{Y}}_{1}\right\},
\end{eqnarray}
for this pp-wave geometry.  The $X_{K}$ and $Y_{I}$ are subject to Dirichlet boundary conditions, while the $\tilde{Y}_{I}$ satisfy Neumann boundary conditions.  Quantizing the open string, we obtain the embedding coordinates
\begin{eqnarray}
&& \!\!\! X_{K}\left(\tau,\sigma\right) = \sum_{n=1}^{\infty}\sqrt{\frac{2}{\omega_{n}}}
\left\{\alpha_{n}^{K}e^{-i\omega_{n}\tau} + \left(\alpha_{n}^{K}\right)^{\dag}e^{i\omega_{n}\tau}\right\}\sin{\left(n\sigma\right)} \\
&& \!\!\! Y_{\pm}\left(\tau,\sigma\right) = \sum_{n=1}^{\infty}\sqrt{\frac{2}{\omega_{n}}}
\left\{\beta^{\mp}_{n}e^{-i\omega^{\mp}_{n}\tau} + \left(\beta^{\pm}_{n}\right)^{\dag}e^{i\omega^{\pm}_{n}\tau}\right\}\sin{\left(n\sigma\right)} \\
\nonumber &&\!\!\! \tilde{Y}_{\pm}\left(\tau,\sigma\right) = \frac{1}{\sqrt{m}}\left\{\tilde{\beta}^{\mp}_{0}e^{-i\omega^{\mp}_{0}\tau} + \left(\tilde{\beta}^{\pm}_{0}\right)^{\dag}e^{i\omega^{\pm}_{0}\tau}\right\} + \sum_{n=1}^{\infty}\sqrt{\frac{2}{\omega_{n}}}
\left\{\tilde{\beta}^{\mp}_{n}e^{-i\omega^{\mp}_{n}\tau} + \left(\tilde{\beta}^{\pm}_{n}\right)^{\dag}e^{i\omega^{\pm}_{n}\tau}\right\}\cos{\left(n\sigma\right)}, \\
\end{eqnarray}
which, as before, satisfy the same commutation relations as in \ref{comm-relns}.  The lightcone Hamiltonian, in terms of the (normal ordered) creation and annihilation operators, is then
\begin{eqnarray}
\nonumber && H_{lc} = \sum_{n=1}^{\infty}\sum_{K=1}^{4}
\frac{\omega_{n}}{m}\left(\alpha_{n}^{K}\right)^{\dag}\alpha_{n}^{K}
+ \sum_{n=1}^{\infty}\left\{\frac{\omega^{+}_{n}}{m}\left(\beta_{n}^{+}\right)^{\dag}\beta_{n}^{+} + \frac{\omega^{-}_{n}}{m}\left(\beta_{n}^{-}\right)^{\dag}\beta_{n}^{-}\right\} \\
&& ~ \hspace{0.85cm} + \sum_{n=0}^{\infty}\left\{\frac{\omega^{+}_{n}}{m}\left(\tilde{\beta}_{n}^{+}\right)^{\dag}\tilde{\beta}_{n}^{+} + \frac{\omega^{-}_{n}}{m}\left(\tilde{\beta}_{n}^{-}\right)^{\dag}\tilde{\beta}_{n}^{-}\right\}
\end{eqnarray}
with spectrum\footnote{Here, unlike in the previous Penrose limit, the $\tilde{\beta}^-$ oscillator actually has a zero mode.}
\begin{eqnarray}
\label{pp-wave hamiltonian 2-alpha}
& \alpha_{n}^{K}: & ~~ \frac{\omega_{n}}{m} = \sqrt{1 + \frac{n^{2}}{m^{2}}}
\hspace{3.0cm} \textrm{with } n = 1, 2, \ldots \hspace{1.8cm} \\
\label{pp-wave hamiltonian 2-beta}
& \beta_{n}^{\pm}: & ~~ \frac{\omega^{\pm}_{n}}{m} = \sqrt{1 + \frac{n^{2}}{m^{2}}} \pm 1
\hspace{2.2cm} \textrm{with } n = 1, 2, \ldots \hspace{1.8cm} \\
\label{pp-wave hamiltonian 2-beta-tilde}
& \tilde{\beta}_{n}^{\pm}: & ~~ \frac{\omega^{\pm}_{n}}{m} = \sqrt{1 + \frac{n^{2}}{m^{2}}} \pm 1
 \hspace{2.2cm} \textrm{with } n = 0, 1, 2, \ldots \hspace{1.8cm}
\end{eqnarray}

Again, we can interpret the above results with the following assumptions:
\begin{equation}
p_{\xi} = -\frac{1}{R^{2}}\left(E + P_{\chi_{1}} + P_{\chi_{+}}\right) = -m
\end{equation}
may not vanish, whereas
\begin{eqnarray}
&& p_{u} = -E + P_{\chi_{1}} + P_{\chi_{+}} = -H_{lc}, \\
\nonumber && p_{y_{1}} = \frac{2}{\sqrt{3}}\frac{1}{R}\tfrac{\left[-\left(\sqrt{1-\alpha_{0}^{2}} - 1\right)P_{\chi_{1}} + \sqrt{1-\alpha_{0}^{2}}P_{\chi_{+}}\right]}{(1-\alpha_{0}^{2})^{1/4}\sqrt{1-\sqrt{1-\alpha_{0}^{2}}}}
~~~~ \textrm{and} ~~~~ p_{\tilde{y}_{1}} = \frac{2}{\sqrt{3}}\frac{1}{R}\tfrac{P_{\chi_{-}}}{(1-\alpha_{0}^{2})^{1/4}\sqrt{1-\sqrt{1-\alpha_{0}^{2}}}}
\end{eqnarray}
must remain finite.  Hence
\begin{equation} \label{charges pp-wave null+}
E = P_{\chi_1} + P_{\chi_+} + O(1), ~~~~~~ J_{\chi_{+}} = \tfrac{1-\sqrt{1-\alpha_{0}^{2}}}{\sqrt{1-\alpha_0^2}} \hspace{0.1cm} P_{\chi_1} + O(R), ~~~~~~ P_{\chi_{-}} = O(R)
\end{equation}
and
$P_{\chi_{1}}$ should be of $O(R^{2})$.


\subsection{Gauge theory and the pp-wave}

In the dual conformal field theory, `near-BPS' strings map to certain holomorphic `words' built out of composite scalar fields, and attaching open strings to D-branes corresponds to attaching these words to the associated Schur polynomial \cite{RdMK-et-al}.  Diagonalizing the resulting conformal dimension corresponds to finding the spectrum of string states, which could, in principle, be compared with \ref{pp-wave hamiltonian 1-alpha} - \ref{pp-wave hamiltonian 1-beta-tilde} and \ref{pp-wave hamiltonian 2-alpha} - \ref{pp-wave hamiltonian 2-beta-tilde}.  Such analyses have had considerable success in the closed string sector \cite{BMN, MZ, MZ-ABJM}, as well as in the open string sector of type IIB string theory on $AdS_{5}\times S^{5}$ \cite{Berenstein-open-strings, Nepomechie, Berenstein-AdS-open-strings}.  Unfortunately, because the Klebanov-Witten gauge theory is non-renormalizable, we have no control over the perturbation theory, and such a comparison is not possible.  However, an extension of this analysis to type IIA string theory on $AdS_{4}\times\mathbb{CP}^{3}$ should prove more fruitful, as the dual ABJM theory is renormalizable \cite{HMP-CP3}.

We can, however, shed some light on the properties of the words dual to these open pp-wave strings - particularly, the way in which their global charges relate to the two different Penrose limits.  Any holomorphic operator dual to propagating degrees of freedom must be made from combinations of the composite scalar fields, $A_{i}B_{j}$, which form a $(\tfrac{1}{2}, \tfrac{1}{2})_{1}$ representation of the global $SU(2)_A \times SU(2)_B \times U(1)_{\mathcal{R}}$ symmetry group.  We can explicitly track the charges by referring to the correspondence given in \ref{par-z1} - \ref{par-z4}.  We make the identification
\begin{eqnarray}
J_A &=& i\partial_{\phi_1} = \frac{1}{3} \left( P_{\chi_1} - 3 P_{\chi_2} +  P_{\chi_3} \right) \\
J_B &=& i\partial_{\phi_2} = \frac{1}{3} \left( P_{\chi_1} + P_{\chi_2} - 3 P_{\chi_3} \right) \\
J_U &=& -2i\partial_\psi = \frac{2}{3} \left( P_{\chi_1} + P_{\chi_2} + P_{\chi_3} \right).
\end{eqnarray}

Let us assume that, for the states of interest, the total length of the operator $J_{U}=L$ (the number of composite scalar fields $A_{i}B_{j}$) is equal to the $U(1)_{\mathcal{R}}$ charge (as would be the case for a holomorphic operator).  If we make this identification, we can gain some understanding of the charges required by such operators (see equations \ref{charges pp-wave null2} and \ref{charges pp-wave null+}).

In the pp-wave of section \ref{section - excitations - null2}, this identification leads to
\begin{equation}
E = \frac{3}{2}L + O(R),
\end{equation}
\begin{equation}
P_{\chi_1} = \frac{3}{4} \left( 2-\alpha_{0}^{2} \right) L + O(R), \hspace{0.75cm}
P_{\chi_2} = \frac{3}{4} \alpha_{0}^{2} L + O(R) \hspace{0.75cm} \textrm{and} \hspace{0.75cm} P_{\chi_3} = O(1), \nonumber
\end{equation}
which gives the two $SU(2)$ charges
\begin{equation}
J_{A} =  \left( \frac{1}{2} - \alpha_0^2 \right) L + O(R)
\hspace{1.0cm} \textrm{and} \hspace{1.0cm}
J_{B} = \frac{L}{2} + O(R).
\end{equation}

In section \ref{section - excitations - null+}, we have
\begin{equation}
E = \frac{3}{2}L + O(R),
\end{equation}
\begin{equation}
J_{\chi_1} = \frac{3}{2} \sqrt{1-\alpha_{0}^{2}} \hspace{0.1cm} L \hspace{0.08cm} + \hspace{0.08cm} O(R), \hspace{0.5cm}
J_{\chi_+} = \frac{3}{2} \left( 1 - \sqrt{1- \alpha_{0}^{2}} \right) L \hspace{0.08cm} + \hspace{0.08cm} O(R) \hspace{0.5cm} \textrm{and}  \hspace{0.5cm} J_{\chi_-} = O(R), \nonumber
\end{equation}
so that
\begin{equation}
J_{A} = J_{B} = \left( \sqrt{1-\alpha_{0}^{2}} - \frac{1}{2} \right)L + O(R).
\end{equation}

These results fit nicely with the interpretation of these states as being, in some sense, `near-BPS'.  We can see that the energy of the states in both pp-wave limits are of the order of the BPS bound $\tfrac{3}{2}L$ \cite{Klebanov-Witten}.  Additionally, the two $SU(2)$ charges are limited to the interval $[-\tfrac{L}{2}, \tfrac{L}{2}]$ as they must be for holomorphic operators.

The dependence of $J_{A}, J_{B}$ on $\alpha_{0}$ arizes from the requirement that the physical momenta of the states be finite, which is presumably 
related to the fact that open strings attached to the giant must stay on its worldvolume.  If these states are picked out as being `near-BPS' by the gauge theory, it could provide some insight into the emergence of the non-trivial geometry of $T^{1,1}$.


\section{Conclusion}
\label{section - conclusion}

We have shown how to translate Mikhailov's elegant construction of giant gravitons in terms of holomorphic curves \cite{Mikhailov} into the more familiar DBI construction for giants on the Einstein-Sasaki space $T^{1,1}$. This solution is dual to the subdeterminant operator $\mathcal{O}_{n}(A_{1}B_{1})$. Its factorization, at maximal size $\alpha_{0}=1$, into the two dibaryons of \cite{BHK}, which are dual to the determinant operators $\det{A_{1}}$ and $\det{B_{1}}$, may be thought of as the disappearance of the map between the polar coordinates of the two 2-spheres in $T^{1,1}$ (which results from the constraint $\sin{\tfrac{\theta_{1}}{2}}\sin{\tfrac{\theta_{2}}{2}} = \sqrt{1-\alpha_{0}^{2}}$ contained in the definition of the giant graviton).  The D3-brane then independently wraps both 2-spheres.

We also presented a general analysis of the spectrum of small fluctutations about the giant graviton. This spectrum was calculated in two special cases: the small giant graviton with $\alpha_{0} \ll 1$ and the maximal giant graviton.  This latter turns out to be the same as the spectrum \cite{BHK} obtained for two separate dibaryons, even taking into account excitations between these two halves of the maximal giant. A comparison between these results indicates that the frequencies are dependent on the size of the giant graviton\footnote{This is not true for the approximate spectrum of the small submaximal giant graviton, when taken on its own.}.  This is curious. This phenomenon has not been observed before - probably since all previous giant gravitons have been spherical or nearly spherical configurations. The physics of why the fluctuation spectrum is independent of the size of the giant is quite clear in the $AdS_{5}$ case. There, the spectrum was trivial since, when the brane's radius is increased, the blueshift of the geometry exactly cancels the increase in wavelength of the modes so that they came out independent of the radius of the brane. We do not yet understand the physics of the $T^{1,1}$ giant's spectrum.

We also attached open strings to the giant graviton. We were able to quantize short open strings in pp-wave geometries associated with two distinct worldvolume null geodesics and obtain their energy spectra.  These open strings ending on the giant graviton should correspond to certain words (composed of combinations $A_{i}B_{j}$ of scalar fields) with size of
$O(\sqrt{N})$ attached to the subdeterminant $\mathcal{O}_{n}(A_{1}B_{1})$.  However, the non-renormalizability of the $\mathcal{N}=1$ Klebanov-Witten gauge theory makes comparison of the anomalous dimensions of these (near-BPS) operators with the corresponding string spectrum problematic. Nevertheless, there are lessons to be learnt from this study.

\begin{itemize}
  \item
  The structural similarities between the Klebanov-Witten and ABJM theories leads us to
  hypothesize that a giant graviton in $AdS_{4}\times \mathbb{CP}^{3}$ can
  be constructed using an only slightly modified ansatz (taking into account, of course, the more
  complicated nature - and the additional radial coordinate - of the complex projective space).

  \item
  The `factorization' of the maximal giant graviton into $\mathbb{CP}^{2}$ dibaryons should
  be qualitatively similar to the $T^{1,1}$ case.

  \item
  ABJM theory has the same superpotential as Klebanov-Witten theory, but, since
  it is a 2+1 dimensional conformal field theory, is renormalizable. Consequently, it should be
  possible, not only to construct the giant graviton in $AdS_{4}\times
  \mathbb{CP}^{3}$, which is extended in the complex projective space, but also to match
  the energies of open string excitations to the anomalous dimensions of nearly
  $\tfrac{1}{2}$-BPS operators.
\end{itemize}
We leave these questions for future studies \cite{HMP-CP3}.

\section{Acknowledgements}

We would like to thank Michael Abbott, Asadig Mohammed and especially Robert de Mello Koch for pleasant discussions and/or helpful correspondence.  We would also like to thank the anonymous referee for useful comments that helped improve the clarity of the presentation.  This work is based upon research supported by the National Research Foundation (NRF) of South Africa's Thuthuka and Key International Scientific Collaboration programs. AP is supported by an NRF Scarce Skills PhD fellowship. AH acknowledges support by a University of Cape Town Vice Chancellor's postdoctoral fellowship. Any opinions, findings and conclusions or recommendations expressed in this material are those of the authors and therefore the NRF do not accept any liability with regard thereto.


\appendix

\section{$T^{1,1}$ metric in the coordinates $(\theta_{i},\chi_{i})$ and $(z_{i},\chi_{i})$} \label{appendix - conifold metric}

The radial part \ref{ds2-rad} of the $T^{1,1}$ metric depends on the polar angles $\theta_{i}$, but can also be written in terms of the alternative radial coordinates $z_{i} \equiv \cos^{2}{\tfrac{\theta_{i}}{2}}$ as follows:
\begin{equation}
ds_{\textrm{radial}}^{2} = \frac{1}{6}\left\{d\theta_{1}^{2} + d\theta_{2}^{2}\right\}
= \frac{1}{6}\left\{\frac{dz_{1}^{2}}{z_{1}(1-z_{1})} + \frac{dz_{2}^{2}}{z_{2}(1-z_{2})}\right\},
\end{equation}
or, in the alternative orthogonal radial coordinates $\alpha$ and $\beta$, where the definition $v \equiv \sin{\beta}$, with $\beta ~ \epsilon ~ [0,\pi]$, allows us to traverse both coverings of the unit interval:
\begin{equation}
ds_{\textrm{radial}}^{2} = g_{\alpha} \hspace{0.1cm} d\alpha^{2} + g_{\beta} \hspace{0.1cm} d\beta^{2},
\end{equation}
with components
\begin{eqnarray}
&& g_{\alpha} = \frac{\alpha^{2}\sin^{2}{\beta}}{3\left(1-\alpha^{2}\right)\sqrt{1-\alpha^{2}\sin^{2}{\beta}} \label{g-alpha}
\left(1-\sqrt{1-\alpha^{2}\sin^{2}{\beta}}\right)} \\
&& g_{\beta} = \frac{\left(1-\sqrt{1-\alpha^{2}\sin^{2}{\beta}}\right)}{3\sin^{2}{\beta}\sqrt{1-\alpha^{2}\sin^{2}{\beta}}}.\label{g-beta}
\end{eqnarray}

The angular part of the metric \ref{ds2-ang}, depends on the phases, defined in \ref{chi1} - \ref{chi3}, via
\begin{equation}
ds_{\textrm{angular}}^{2} = \sum_{i,j}\left(g_{\chi}\right)_{ij} \hspace{0.1cm} d\chi_{i} d\chi_{j},
\end{equation}
where the components are
\begin{eqnarray}
\nonumber & \left(g_{\chi}\right)_{11} & = \tfrac{1}{32}\left[2\left(2 - \cos{\theta_{1}} - \cos{\theta_{2}}\right)^{2} + 3\sin^{2}{\theta_{1}} + 3\sin^{2}{\theta_{2}}\right] \\
&& = \tfrac{1}{8}\left[2\left(2-z_{1}-z_{2}\right)^{2} + 3z_{1}\left(1-z_{1}\right) + 3z_{2}\left(1-z_{2}\right)\right]  \\
& \left(g_{\chi}\right)_{22} & = \tfrac{1}{32}\left[2\left(1+\cos{\theta_{1}}\right)^{2} + 3\sin^{2}{\theta_{1}}\right]
= \tfrac{1}{8}z_{1}\left(3-z_{1}\right) \\
& \left(g_{\chi}\right)_{33} & = \tfrac{1}{32}\left[2\left(1+\cos{\theta_{2}}\right)^{2} + 3\sin^{2}{\theta_{2}}\right]
= \tfrac{1}{8}z_{2}\left(3-z_{2}\right) \\
\nonumber & \left(g_{\chi}\right)_{12} & = \tfrac{1}{32}\left[2\left(2 - \cos{\theta_{1}} - \cos{\theta_{2}}\right)\left(1+\cos{\theta_{1}}\right) - 3\sin^{2}{\theta_{1}}\right] \\
&& = \tfrac{1}{8}z_{1}\left(1 + z_{1} - 2z_{2}\right)
\end{eqnarray}
\begin{eqnarray}
\nonumber & \left(g_{\chi}\right)_{13} & = \tfrac{1}{32}\left[2\left(2 - \cos{\theta_{1}} - \cos{\theta_{2}}\right)\left(1+\cos{\theta_{2}}\right) - 3\sin^{2}{\theta_{2}}\right] \\
&& = \tfrac{1}{8}z_{2}\left(1 - 2z_{1} + z_{2}\right) \\
& \left(g_{\chi}\right)_{23} & = \tfrac{1}{16}\left(1+\cos{\theta_{1}}\right)\left(1+\cos{\theta_{2}}\right) = \tfrac{1}{4}z_{1}z_{2}
\end{eqnarray}
and the determinant is then given by
\begin{equation}
\det{g_{\chi}} = \left(\tfrac{3}{32}\right)^{2}\sin^{2}{\theta_{1}}\sin^{2}{\theta_{2}} = \tfrac{9}{64}z_{1}z_{2}\left(1-z_{1}\right)\left(1-z_{2}\right),
\end{equation}
in terms of either $\theta_{i}$ or $z_{i}$.

We shall also require expressions for various combinations of the cofactors of this angular metric as follows:
\begin{eqnarray}
& \left(C_{\chi}\right)_{11} & = \tfrac{3}{64}z_{1}z_{2}\left(3 - z_{1} - z_{2} - z_{1}z_{2}\right) \\
& \left(C_{\chi}\right)_{12} & = -\tfrac{3}{64}z_{1}z_{2}\left(1 + z_{1} - 3z_{2} + z_{1}z_{2}\right) \\
& \left(C_{\chi}\right)_{13} & = -\tfrac{3}{64}z_{1}z_{2}\left(1 - 3z_{1} + z_{2} + z_{1}z_{2}\right) \\
& \left(C_{\chi}\right)_{22} - \left(g_{\chi}\right)_{33} & = -\tfrac{3}{64}z_{1}z_{2}\left(5 + z_{1} - 7z_{2} + z_{1}z_{2}\right) - \tfrac{1}{4}z_{2}^{2} \\
& \left(C_{\chi}\right)_{33} - \left(g_{\chi}\right)_{22} & = -\tfrac{3}{64}z_{1}z_{2}\left(5 - 7z_{2} + z_{2} + z_{1}z_{2}\right) - \tfrac{1}{4}z_{1}^{2} \\
& \left(C_{\chi}\right)_{23} + \left(g_{\chi}\right)_{23} & = \tfrac{1}{64}z_{1}z_{2}\left(1 + 9z_{1} + 9z_{2} - 11z_{1}z_{2}\right)
\end{eqnarray}
and
\begin{equation}
\left(C_{\chi}\right)_{11} - \det{g_{\chi}} = \tfrac{3}{32}z_{1}z_{2}\left(z_{1} + z_{2} - 2z_{1}z_{2}\right),
\end{equation}
in terms of the $z_{i}$.


\section{General Fluctuation Analysis} \label{appendix - general fluctuations}

Let us consider the general fluctuation ansatz
\begin{equation} \label{fluctuation ansatz}
v_{k} = \varepsilon \delta v_{k}(\sigma^{a}), ~~~~~~ \alpha = \alpha_{0} + \varepsilon\delta\alpha(\sigma^{a}) ~~~~~~ \textrm{and} ~~~~~~
\chi_{1} = t + \varepsilon\delta\chi_{1}(\sigma^{a}),
\end{equation}
with worldvolume coordinates $\sigma^{a} = \left(t,\beta,\chi_{2},\chi_{3}\right)$. Here $\alpha_{0}$ is the size of the giant graviton, about which we are perturbing, and $\varepsilon$ is a small parameter.

The D3-brane action, to second order in $\varepsilon$, takes the form
\begin{eqnarray} \label{action - second order general fluct}
& \! S & \approx T_{3}R^{4} \int dt ~ d\beta ~ d\chi_{2} ~ d\chi_{3} ~~ \sqrt{g_{\beta}\left[\left(C_{\chi}\right)_{11} - \det{g_{\chi}}\right]} \\
\nonumber && \times \left\{\frac{\varepsilon}{\left[\left(C_{\chi}\right)_{11} - \det{g_{\chi}}\right]} \left[\left(C_{\chi}\right)_{11}\dot{\delta\chi}_{1} +
\left(C_{\chi}\right)_{12}\left(\p_{\chi_{2}}\delta\chi_{1}\right) + \left(C_{\chi}\right)_{13}\left(\p_{\chi_{3}}\delta\chi_{1}\right)\right] \right. \\
\nonumber && \left. ~~~
+ \frac{\varepsilon^{2}}{2}\left[\sum_{k}\left\{-\frac{\left(C_{\chi}\right)_{11}\delta v_{k}^{2}}{\left[\left(C_{\chi}\right)_{11} - \det{g_{\chi}}\right]} - \left(\p~\delta v_{k}\right)^{2}\right\}
- g_{\alpha}\left(\p~\delta\alpha\right)^{2}
- \frac{\left(\det{g_{\chi}}\right)\left(\p~\delta \chi_{1}\right)^{2}}
{\left[\left(C_{\chi}\right)_{11} - \left(\det{g_{\chi}}\right)\right]}\right]\right\},
\end{eqnarray}
where the terms involving a single $\varepsilon$ coefficient should be expanded\footnote{Note that the zeroth order terms in these expansions - corresponding to $O(\varepsilon)$ terms in the D3-brane action - yield total derivatives.  This is to be expected, as the giant graviton is a solution of the equations of motion.} to first order in $\varepsilon\delta\alpha$, thus contributing addition second order terms to the D3-brane action.  The components $g_{\alpha}$ and $g_{\beta}$ of the radial metric in the alternative orthogonal radial coordinates $\alpha$ and $\beta$ are given in \ref{g-alpha} and \ref{g-beta}. The expressions involving the cofactors $(C_{\chi})_{ij}$ of the angular metric components $(g_{\chi})_{ij}$ are stated explicitly in appendix \ref{appendix - conifold metric}.

The (spacetime) gradient squared on the worldvolume of the giant graviton is defined as follows:
\begin{eqnarray}
&& \!\! \left(\p f\right)^{2} \equiv  \frac{\left(\p_{\beta}f\right)^{2}}{g_{\beta}}
- \frac{1}{\left[\left(C_{\chi}\right)_{11} - \det{g_{\chi}}\right]} \left\{\left(C_{\chi}\right)_{11}\dot{f}^{2} + \left[\left(C_{\chi}\right)_{22}-\left(g_{\chi}\right)_{33}\right]\left(\p_{\chi_{2}}f\right)^{2} \right. \\
\nonumber && \hspace{4.425cm} + \left[\left(C_{\chi}\right)_{33}-\left(g_{\chi}\right)_{22}\right]\left(\p_{\chi_{3}}f\right)^{2} + 2\left(C_{\chi}\right)_{12}\dot{f}\left(\p_{\chi_{2}}f\right) \\
\nonumber && \left. \hspace{4.4cm} + 2\left(C_{\chi}\right)_{13}\dot{f}\left(\p_{\chi_{3}}f\right) +
2\left[\left(C_{\chi}\right)_{23}+\left(g_{\chi}\right)_{23}\right]\left(\p_{\chi_{2}}f\right)\left(\p_{\chi_{3}}f\right)\right\}, ~~~~
\end{eqnarray}
with $f(t,\beta,\chi_{2},\chi_{3})$ any function of the worldvolume coordinates.  Here all the cofactors $(C_{\chi})_{ij}$, and metric components $(g_{\chi})_{ij}$ and $g_{\beta}$, as well as the determinant $\det{g_{\chi}}$ are evaluated at $\alpha = \alpha_{0}$.


\section{Eigenvalue problems} \label{appendix - evp}

\subsection{Eigenvalue problem for the small submaximal giant graviton} \label{appendix - evp - small giant}

 Let us consider the eigenvalue problem $\tilde{\Box}\Psi = -\lambda\Psi$, with $\tilde{\Box}$ the (rescaled) d'Alembertian operator \ref{dAlem} on the worldvolume of the small submaximal giant graviton.

If we take an ansatz
\begin{equation}
\Psi(t,z,\chi_{2},\chi_{3}) \equiv f(z) ~ e^{-i\omega t} ~ e^{\frac{3}{4}im\chi_{2}} ~ e^{\frac{3}{4}in\chi_{3}}, ~~~~~~ \textrm{with} ~~ m,n ~ \epsilon ~ \mathbb{Z},
\end{equation}
then the problem reduces to solving the ordinary differential equation
\begin{eqnarray}
\nonumber && \! \p_{z}\left\{z\left(1-z\right)\p_{z}f(z)\right\} \\
&& \! -\tfrac{1}{4}\left\{m^{2} + n^{2} - \left[\omega + \tfrac{1}{4}\left(m+n\right)\right]^{2} + m^{2}~\tfrac{z}{\left(1-z\right)} + n^{2}~\tfrac{\left(1-z\right)}{z} - \tfrac{2}{3}\lambda\right\}f(z) = 0. ~~~~~~~~
\end{eqnarray}
Now setting
\begin{equation}
f(z) \equiv z^{\frac{1}{2}|n|}\left(1-z\right)^{\frac{1}{2}|m|} h(z),
\end{equation}
we obtain the hypergeometric differential equation
\begin{eqnarray}
\nonumber && z\left(1-z\right) \p_{z}^{2}h(z) + \left[\left(|n|+1\right) - \left(|m|+|n|+2\right)\right]\p_{z}h(z) \\
&& -\tfrac{1}{4}\left\{2\left(|mn| + |m| + |n|\right) + m^{2} + n^{2} - \left[\omega + \tfrac{1}{4}\left(m+n\right)\right]^{2} - \tfrac{2}{3}\lambda\right\}h(z) = 0. ~~~~~~~~
\end{eqnarray}
Solutions $h(z) = F(a,b,c;z)$ are hypergeometric functions dependent on the following parameters:
\begin{eqnarray}
&& a \equiv \tfrac{1}{2}\left(|m|+|n|+1\right) \mp \sqrt{\tfrac{1}{6}\lambda + \tfrac{1}{4}\left[\omega + \tfrac{1}{4}\left(m+n\right)\right]^{2} + \tfrac{1}{4}} \\
&& b \equiv \tfrac{1}{2}\left(|m|+|n|+1\right) \mp \sqrt{\tfrac{1}{6}\lambda + \tfrac{1}{4}\left[\omega + \tfrac{1}{4}\left(m+n\right)\right]^{2} + \tfrac{1}{4}} \\
&& c \equiv |n|+1,
\end{eqnarray}
where, for regularity at $z=1$, either $a$ or $b$ must be a negative integer \cite{Gubser} (whichever corresponds to the negative in front of the square root).  Hence
\begin{equation}
\tfrac{1}{2}\left(|m|+|n|+1\right) - \sqrt{\tfrac{1}{6}\lambda + \tfrac{1}{4}\left[\omega + \tfrac{1}{4}\left(m+n\right)\right]^{2} + \tfrac{1}{4}} \equiv -s,
~~~~~ \textrm{with} ~~ s = 0,1,2,\ldots
\end{equation}

The eigenfunctions
\begin{equation} \label{eigenfunctions - small giant}
\Psi_{smn}(t,z,\chi_{2},\chi_{3}) \equiv z^{\frac{1}{2}|n|}\left(1-z\right)^{\frac{1}{2}|m|} F_{smn}(z) ~ e^{-i\omega t} ~ e^{\frac{3}{4}im\chi_{2}}e^{\frac{3}{4}in\chi_{3}}
\end{equation}
therefore correspond to the eigenvalues
\begin{equation}
\lambda_{smn}(\omega) = 6l(l+1) - \tfrac{3}{2}\left[\omega + \tfrac{1}{4}(m+n)\right]^{2}, ~~~~~~ \textrm{where} ~~~
l \equiv  s + \max{\left\{\tfrac{1}{2}|m+n|,\tfrac{1}{2}|m-n|\right\}}.
\end{equation}
Our hypergeometric functions $F_{smn}(z) \equiv F(a,b,c;z)$ are dependent (through $a$, $b$ and $c$) on the integers $s$, $m$ and $n$, with $s$ non-negative.


\subsection{Stationary eigenvalue problem for a dibaryon} \label{appendix - evp - dibaryon}

\textbf{\emph{Standard Eigenfunctions}}

We would now like to solve the stationary eigenvalue problem $\nabla^{2}\Phi = -E\Phi$, with $\nabla^{2}$ the Laplacian \ref{Laplacian} on the spatial extension of a dibaryon.  Setting
\begin{equation}
\Phi(z,\xi,\phi) = f(z) ~ e^{im\xi} ~ e^{in\phi}, ~~~~~~~~ \textrm{where} ~~ m ~ \epsilon ~ \tfrac{1}{2}\mathbb{Z} ~~ \textrm{and} ~~ n ~ \epsilon ~ \mathbb{Z},
\end{equation}
we obtain the following ordinary differential equation
\begin{eqnarray}
\nonumber && \p_{z}\left\{z\left(1-z\right)\p_{z}f(z)\right\} \\
&& - \left\{\tfrac{1}{4}(m+n)^{2}~\tfrac{(1-z)}{z} + \tfrac{1}{4}(m-n)^{2}~\tfrac{z}{(1-z)} + \tfrac{1}{2}(2m^{2}+n^{2}) - \tfrac{1}{6}E\right\}f(z) = 0. \hspace{1.0cm}
\end{eqnarray}
Let us now define
\begin{equation}
f(z) = z^{\frac{1}{2}|m+n|}\left(1-z\right)^{\frac{1}{2}|m-n|}h(z),
\end{equation}
whereupon the differential equations reduces to the form
\begin{eqnarray}
\nonumber && z\left(1-z\right)\p_{z}^{2}h(z)+\left[\left(|m+n|+1\right) - \left(|m+n| + |m-n| + 2\right)z\right]\p_{z}h(z) \\
&& -\left\{\tfrac{1}{2}|m^{2}-n^{2}| + \tfrac{1}{2}|m+n| + \tfrac{1}{2}|m-n| + \tfrac{1}{2}(2m^{2}+n^{2}) - \tfrac{1}{6}E\right\}h(z) = 0. \hspace{1.0cm}
\end{eqnarray}
This is a hypergeometric differential equation with solutions $h(z) = F(a,b,c;z)$, where we define
\begin{eqnarray}
&& a \equiv \tfrac{1}{2}\left(|m+n| + \tfrac{1}{2}|m-n| + 1\right) \pm \sqrt{\tfrac{1}{6}E - \tfrac{1}{2}m^{2} + \tfrac{1}{4}} \\
&& b \equiv \tfrac{1}{2}\left(|m+n| + \tfrac{1}{2}|m-n| + 1\right) \mp \sqrt{\tfrac{1}{6}E - \tfrac{1}{2}m^{2} + \tfrac{1}{4}} \\
&& c \equiv |m+n| + 1.
\end{eqnarray}
For regularity at $z=1$, either $a$ or $b$ should be a negative integer \cite{Gubser}.  We therefore require that
\begin{equation}
\tfrac{1}{2}\left(|m+n| + \tfrac{1}{2}|m-n| + 1\right) - \sqrt{\tfrac{1}{6}E - \tfrac{1}{2}m^{2} + \tfrac{1}{4}} \equiv -s,
~~~~~~ \textrm{with} ~~ s = 0,1,2,\ldots
\end{equation}

Hence
\begin{equation} \label{eigenfunctions - dibaryon}
\Phi_{smn}(z,\xi,\phi) = z^{\frac{1}{2}|m+n|}\left(1-z\right)^{\frac{1}{2}|m-n|}F_{smn}(z) ~ e^{im\xi} ~ e^{in\phi}
\end{equation}
are the eigenfunctions corresponding to the eigenvalues
\begin{equation}
E_{smn} = 6l(l+1) + 3m^{2}, ~~~~~~~~ \textrm{where} ~~~ l \equiv s + \max{\left\{|m|,|n|\right\}},
\end{equation}
with $s$ and $n$ integers ($s$ is non-negative), and $m$ an integer or half-integer.  Here $F_{smn}(z) \equiv F(a,b,c;z)$ are the hypergeometric functions previously described.

\bigskip

\textbf{\emph{Modified Eigenfunctions}}

We should now look for additional solutions to the stationary eigenvalue problem $\nabla^{2}\Phi = -E\Phi$, associated with a dibaryon, such that $\Phi \sim (1-z)^{-\frac{1}{2}}$ as $z \rightarrow 1$.

Hence we shall choose
\begin{equation}
f(z) = z^{\frac{1}{2}|m+n|}(1-z)^{-\frac{1}{2}} ~ h(z), ~~~~~~~~ \textrm{with} ~~~ n = m \pm 1,
\end{equation}
which again yields a hypergeometric differential equation
\begin{equation}
z\left(1-z\right)\p_{z}^{2}h(z)+\left(|m+n|+1\right)\left(1-z\right)\p_{z}h(z)
-\left\{-\tfrac{1}{2} + \tfrac{1}{2}(2m^{2}+n^{2}) - \tfrac{1}{6}E\right\}h(z) = 0.
\end{equation}
The hypergeometric solutions $h(z) = F(a,b,c;z)$ can be written in terms of the parameters
\begin{eqnarray}
&& a \equiv \tfrac{1}{2}|m+n| \pm \sqrt{\tfrac{1}{6}E - \tfrac{1}{2}m^{2} + \tfrac{1}{4}} \\
&& b \equiv \tfrac{1}{2}|m+n| \mp \sqrt{\tfrac{1}{6}E - \tfrac{1}{2}m^{2} + \tfrac{1}{4}} \\
&& c \equiv |m+n| + 1,
\end{eqnarray}
with $a$ or $b$ a negative integer \cite{Gubser}. Hence
\begin{equation}
\tfrac{1}{2}|m+n| - \sqrt{\tfrac{1}{6}E - \tfrac{1}{2}m^{2} + \tfrac{1}{4}} = -s, ~~~~~~~~ \textrm{with} ~~ s = 0,1,2,\ldots
\end{equation}

The modified eigenfunctions then take the form
\begin{equation} \label{eigenfunction - modified dibaryon}
\Phi^{\textrm{mod}}_{smn}(z,\xi,\phi) = z^{\frac{1}{2}|m+n|}\left(1-z\right)^{-\frac{1}{2}} F^{\textrm{mod}}_{smn}(z) ~ e^{im\xi} ~ e^{in\phi},
\end{equation}
with corresponding eigenvalues
\begin{equation}
E^{\textrm{mod}}_{smn} = 6l^{\textrm{mod}}(l^{\textrm{mod}}+1) + 3m^{2}, ~~~~~~~~ \textrm{where} ~~ l^{\textrm{mod}} \equiv s + \tfrac{1}{2}\left(|m+n| - 1\right),
\end{equation}
and $s$, $m$ and $n$ are integers, with $s$ non-negative and $n = m\pm 1$.  Here also $F^{\textrm{mod}}_{smn}(z) \equiv F(a,b,c;z)$ is our hypergeometric solution.


\end{document}